\documentclass[lettersize,journal]{IEEEtran}
\usepackage{amsmath,amsfonts}
\usepackage{algorithmic}
\usepackage{algorithm}
\usepackage{array}
\usepackage[caption=false,font=normalsize,labelfont=sf,textfont=sf]{subfig}
\usepackage{textcomp}
\usepackage{stfloats}
\usepackage{url}
\usepackage{verbatim}
\usepackage{graphicx}
\usepackage{cite}
\newtheorem{problem}{Problem}

\newtheorem{proposition}{Proposition}
\newtheorem{theorem}{Theorem}

\usepackage{accents}
\newcommand{\ubar}[1]{\underaccent{\bar}{#1}}
\usepackage{color,soul}

\begin{document}

\title{Combining Cooperative Re-Routing with Intersection Coordination for Connected and Automated Vehicles in Urban Networks}

\author{Panagiotis Typaldos$^{1}$ and Andreas A. Malikopoulos$^{2}$, \textit{Senior Member, IEEE}
        % <-this % stops a space
\thanks{This research was supported in part by NSF under Grants CNS-2401007, CMMI-2219761, IIS-2415478, and in part by MathWorks.}% <-this % stops a space
% \thanks{The authors are with the School of Civil and Environmental Engineering, Cornell University, Ithaca, NY 14850, USA. {\tt\small email: \{pt432,amaliko\}@cornell.edu}}}
\thanks{$^{1}$Panagiotis Typaldos is with the School of Civil and Environmental Engineering, Cornell University, Ithaca, New York State, United States {\tt\small pt432@cornell.edu}}%
\thanks{$^{2}$Andreas A. Malikopoulos is with the School of Civil and Environmental Engineering and the System Engineering Program, Cornell University, Ithaca, New York State, United States {\tt\small amaliko@cornell.edu}}}

\markboth{Journal of \LaTeX\ Class Files,~Vol.~14, No.~8, August~2021}%
{Shell \MakeLowercase{\textit{et al.}}: A Sample Article Using IEEEtran.cls for IEEE Journals}


\maketitle

\begin{abstract}
In this paper, we present a hierarchical framework that integrates upper-level routing with low-level optimal trajectory planning for connected and automated vehicles (CAVs) traveling in an urban network. The upper-level controller efficiently distributes traffic flows by utilizing a dynamic re-routing algorithm that leverages real-time density information and the fundamental diagrams of each network edge. This re-routing approach predicts when each edge will reach critical density and proactively adjusts the routing algorithm’s weights to prevent congestion before it occurs. The low-level controller coordinates CAVs as they cross signal-free intersections, generating optimal, fuel-efficient trajectories while ensuring safe passage by satisfying all relevant constraints. We formulate the problem as an optimal control problem and derive an analytical solution. Using the SUMO micro-simulation platform, we conduct simulation experiments on a realistic network. The results show that our hierarchical framework significantly enhances network performance compared to a baseline static routing approach. By dynamically re-routing vehicles, our approach successfully reduces total travel time and mitigates congestion before it develops.
\end{abstract}


\section{Introduction}
\IEEEPARstart{O}{ver} the past decade, the rapidly increasing population and the urbanization, have led to a significant rise in traffic congestion in urban areas, as travel demand increases \cite{worldUrban2018, schrank2015urban}. Connected and automated vehicles (CAVs) have the potential to substantially improve fuel efficiency, increase road safety, and optimize traffic flow \cite{zhao2019enhanced,sjoberg2017cooperative, diakaki2015overview}. However, due to the complex nature of urban networks and road traffic, challenges arise on how to develop sophisticated approaches for CAVs' decision-making. To address these challenges, several research efforts have focused on developing efficient routing planning algorithms and intersection coordination approaches for CAVs.

\subsection{Related Work}

\subsubsection{Intersection Coordination}

Intersections are a crucial part of urban networks, ensuring the safe crossing of vehicles via traffic signals \cite{diakaki2015overview, papageorgiou2007its}. On the other hand, they are also responsible for the majority of vehicle stops, which leads to increased emissions and fuel consumption. As it is expected that CAVs will gradually penetrate the market, a shift toward signal-free intersection operation has been noticed in literature, which is expected to lead to a reduction of vehicle stops and subsequently to increased throughput \cite{chen2015cooperative, namazi2019intelligent, wu2022intersection}. The majority of the proposed approaches consider pre-specified crossing paths for the vehicles and pre-calculate the conflict points among these paths. Then, they specify a conflict-free crossing solution for all vehicles based on different strategies, such as first-in-first-out (FIFO) or optimization-based crossing sequence, guiding vehicles such that they never pass a common conflict point simultaneously \cite{Bang2022combined, naderi2024optimal}. 

Several studies have explored decentralized optimal control CAVs navigating signal-free urban intersections. Malikopoulos et al. \cite{malikopoulos2018decentralized,Malikopoulos2020,mahbub2020Automatica-2} developed a decentralized constrained optimal control framework aimed at jointly minimizing energy consumption and maximizing traffic throughput while ensuring safety. The authors derived an analytical solution for the decentralized problem and established conditions that guarantee the existence of feasible solutions that always satisfy safety constraints. Xu et al. \cite{xu2022general} proposed a decentralized control framework for CAVs at multi-lane signal-free intersections. Their approach introduced discrete merging points to replace conventional merging zones, integrating a control barrier function controller to enforce safety constraints while tracking optimal trajectories. Jiang et al. \cite{jiang2022coordination} focused on mixed traffic environments, presenting a two-level optimization framework for coordinating CAVs and human-driven vehicles at signal-free intersections. Their method combined platoon coordination and dynamic priority control at the upper level with an eco-driving strategy at the lower level to minimize total delay and energy consumption.

Xu et al. \cite{xu2021comparison} conducted a comparative analysis of four cooperative driving strategies for CAVs at signal-free intersections: FIFO, a modified FIFO strategy, dynamic resequencing, and Monte Carlo tree search. They evaluated these strategies based on travel time, energy consumption, and computational efficiency. In another study, Naderi et al. \cite{naderi2025lane} proposed a joint optimal control approach for CAVs to navigate intersections without traffic signals or lane constraints. They formulated an optimal control problem that optimizes vehicle acceleration and steering inputs over a time horizon using a dynamic bicycle model. Their cost function incorporated objectives for smooth, collision-free motion while considering fuel efficiency and desired speed tracking.

\subsubsection{Routing Planning}

Routing planning algorithms aim at determining the most efficient path from an origin to a destination, considering factors such as traffic congestion and travel times. For the solution of the routing planning problem, mainly three categories of algorithms are used in the related literature, i.e., optimal, heuristic, and hybrid algorithms \cite{nha2012comparative}. Optimal algorithms guarantee to find the global optimal solution through the exploration of the whole set of available solutions; examples of such methods include the Dijkstra algorithm and incremental graphs \cite{bast2016route, nha2012comparative}. On the other hand, heuristic algorithms strive to find a sub-optimal solution by exploring a subset of the solution space, examples include A*, genetic, and nature-inspired optimization algorithms \cite{russell2016artificial, hajlaoui2016survey}. Finally, hybrid algorithms take advantage of the strengths of both previous approaches: An example of a hybrid method could apply a combination of Dijkstra and genetic algorithms \cite{soltani2002path}.

To the best of our knowledge, there have been limited research efforts addressing the routing planning problem by considering current macroscopic traffic variables, such as density, flows, and mean speed of the traffic. Using these aggregated variables is beneficial as they capture the core relationships between traffic parameters through fundamental diagrams, which relate density to flow and speed. The obtained macroscopic measures (e.g., from traffic sensors) provide a smooth representation of the network by averaging individual vehicle behaviors, thereby reducing the computational complexity compared to microscopic models. This high-level view enables rapid, real-time re-routing decisions, which is critical in large-scale urban networks where traffic conditions evolve rapidly. Additionally, only a few of them consider the optimization of CAVs' trajectory in combination with the routing problem. 

Li et al. \cite{li2021joint} proposed a fuel-aware routing for autonomous vehicles that combines fuel consumption, trip delay, and refueling cost optimization through a unified topology representation while using a forward-looking window-based online algorithm to handle dynamic routing decisions. However, their approach focuses only on the routing problem without considering the vehicle coordination and trajectory planning required at intersections. Moreover, their method assumes fixed refueling policies at stations and does not account for real-time traffic conditions. Recent work by Ho et al. \cite{ho2023collaborative} implemented a collaborative vehicle rerouting system that uses dynamic vehicle selection based on travel time and road capacity to mitigate traffic congestion in urban networks. While their approach effectively addresses congestion through vehicle rerouting, it also focuses only on upper-level traffic management without considering lower-level vehicle coordination at intersections. Moreover, their method relies on predetermined time intervals for rerouting decisions and requires manually tuned congestion thresholds. Tseng and Ferng \cite{tseng2021improved} developed a traffic rerouting strategy using real-time traffic information and dynamic weights based on road density, capacity, and speed to mitigate urban congestion. While their approach effectively reduces travel times through network-level routing, it operates in fixed time intervals and does not address vehicle coordination at intersections. Vitale et al. \cite{vitale2024cooperative} proposed a cost function, which was utilized in the Dijkstra algorithm for the rerouting of CAVs. The proposed cost function considers both the current travel times and the density of each edge of the network. Bang et al. \cite{bang2023optimal} derived the optimal vehicle flow of an urban network, which minimizes the total travel time. Subsequently, a heuristic algorithm is used for the assignment of the routes and the travel times of CAVs at each edge, while a coordination method was also proposed for the CAVs crossing the network intersections. However, this work does not account for real-time changes in the traffic conditions.

\subsection{Contributions}
In the paper, we present a framework that combines the routing planning (upper level) with the optimal trajectory planning (low level) of CAVs traveling in an urban network. The upper-level control consists of a dynamic re-routing algorithm that efficiently distributes the flows in the network based on a predictive approach. Specifically, we derive a method to estimate the time to reach the critical density of each edge in the network, using real-time measurements of density and the corresponding fundamental diagrams. To achieve this, we consider that each intersection can receive the information of the density from the surrounding edges, and it embeds the knowledge of the corresponding fundamental diagrams (FDs), which gives the relationship between density and flow of each edge. Subsequently, based on this prediction, we dynamically adjust the weights in the routing algorithm, enabling proactive re-routing before congestion occurs. The edge weights are calculated based on a dynamic weight function whose value increases as we approach the critical density, allowing for more efficient distribution of traffic flows across the network. 
For the low-level controller, we utilize the controller from \cite{chalaki2021RobustGP}, which presents a robust coordination problem for CAVs crossing a signal-free intersection. 
However, under rare occasions, this approach cannot produce a feasible solution that satisfies all the constraints. To this end, in the current work, we proposed an extended solution of \cite{chalaki2021RobustGP}, where we identify the cases in which no feasible solution can be found, and we solve an enhanced constrained optimal control problem (OCP) that guarantees to always deliver a solution.
 
The effectiveness of this hierarchical framework is demonstrated through simulation results in a realistic network. Specifically, the results indicate that the proposed approach outperforms a baseline approach by (i) minimizing the total travel time of the network and by (ii) successfully reducing congestion due to the predictive capabilities, which enable better redistribution of the traffic before edges reach critical density.

The main contributions of this paper are:
\begin{itemize}
    \item Development of a hierarchical framework that combines the upper-level routing problem with the low-level trajectory planning of CAVs.
    \item Introduction of a predictive re-routing approach based on macroscopic traffic variables and fundamental diagrams.
    \item Design of a dynamic weight function for proactive congestion prevention.
    \item Theoretical analysis of the re-routing algorithm, where the predictive capabilities of the proposed method are validated. We quantify the impact of density measurement errors on prediction accuracy and establish bounds for the time to reach critical density, ensuring robust, proactive congestion mitigation.
    \item Demonstration that the proposed approach significantly improves the performance of road networks.
\end{itemize}

The remainder of the letter is organized as follows. In Section \ref{section:routing_problem}, we introduce the routing problem and the re-routing procedure based on the FDs. Then, we present the OCP formulation of the low-level controller and the analytical solution in Section \ref{section:ICP}. In Section \ref{sec:simulations}, we discuss the simulation experiment and results. Finally, in Section \ref{sec:conclusions}, we draw concluding remarks.

\section{Routing Problem} \label{section:routing_problem}
We consider an urban road network represented as a directed graph $\mathcal{G} = (\mathcal{V}, \mathcal{E})$, where $\mathcal{V}$ is the set of nodes and $\mathcal{E}$ consists of the edges connecting the nodes. Each node $ v \in \mathcal{V}$ represents an intersection in the network, while each edge $e \in \mathcal{E}$ is a road connecting two subsequent nodes. We assume that each intersection includes a coordinator that acts as a database. Specifically, it can store information regarding the infrastructure's geometry, the CAVs' trajectories, and the traffic conditions of the surrounding edges, e.g., density and travel times. We also consider a routing decision unit (RDU) on the network level, which is responsible for delivering optimal routes to each CAV. The RDU communicates with the coordinators to receive real-time information regarding the traffic conditions. Each time a request for a route is sent, the RDU solves the routing planning problem based on the latest available traffic information and distributes the optimal route to the corresponding CAV. For deriving the optimal routes, the Dijkstra algorithm is used for weighted graphs \cite{dijkstra2022note}. Dijkstra algorithm (see Algorithm 1) receives as input the origin and the destination of the CAV's route, as well as the weights of each edge of the graph, where the weights correspond to the travel times of each edge.

\begin{algorithm}[ht]
\label{alg: Dijk}
    \caption{Dijkstra's Algorithm for Minimizing Travel Time}
    \begin{algorithmic}[1]
    \REQUIRE Graph $\mathcal{G}=(\mathcal{V},\mathcal{E})$, source node $s$, travel time function $w(u,v)$
    \ENSURE Minimum travel time paths from $s$ to all nodes
    \STATE $time[s] \leftarrow 0$
    \STATE $time[v] \leftarrow \infty$ for all $v \in V \setminus \{s\}$
    \STATE $Q \leftarrow V$ \COMMENT{Priority queue of vertices}
    \STATE $prev[v] \leftarrow$ null for all $v \in V$
    \WHILE{$Q$ is not empty}
        \STATE $u \leftarrow$ vertex in $Q$ with minimum $time[u]$
        \STATE Remove $u$ from $Q$
        \FOR{each neighbor $v$ of $u$}
            \STATE $newTime \leftarrow time[u] + w(u,v)$ \COMMENT{$w(u,v)$ is travel time on edge $(u,v)$}
            \IF{$newTime < time[v]$}
                \STATE $time[v] \leftarrow newTime$
                \STATE $prev[v] \leftarrow u$
            \ENDIF
        \ENDFOR
    \ENDWHILE
    \RETURN $time$, $prev$ \COMMENT{Returns travel times and predecessor vertices}
    \end{algorithmic}
    \end{algorithm}

In our routing framework, the initial weight of each edge is set based on the nominal (free-flow) travel time, calculated using baseline speed assumptions. However, we do not assume a constant speed for all vehicles; instead, real-time traffic sensor data is used to update these weights dynamically. Specifically, as congestion develops, reflected by increasing density, the travel time on an edge changes. Our algorithm accounts for this by periodically recalculating a dynamic weight, $w_e$, using predictive measures (e.g., the time to reach critical density, $t_c$) and a predefined scaling factor, ensuring that the weights adapt to current traffic conditions.

\subsection{Network Information and Fundamental Diagrams}\label{sec: fds}
To capture the traffic conditions of the network, each coordinator is assumed to be equipped with sensors that can measure the density (number of vehicles) and average speed of the CAVs in real time. 
This information is shared with the RDU, which also has additional information regarding the capacity volume and density and the jam density. The latter can be derived from the fundamental diagram (FD) of each edge. FDs can be either calculated and uploaded offline to the RDU or estimated online (e.g., \cite{tajdari2023online}).

For the FD, we assume a triangular shape, where the relationship between density $k$, flow $q$, and the mean speed $v$, is described as 
\begin{equation}
    q = k\cdot v.
\end{equation}

\begin{figure}[t!]
    \centering
    \includegraphics[width = 0.9\linewidth]{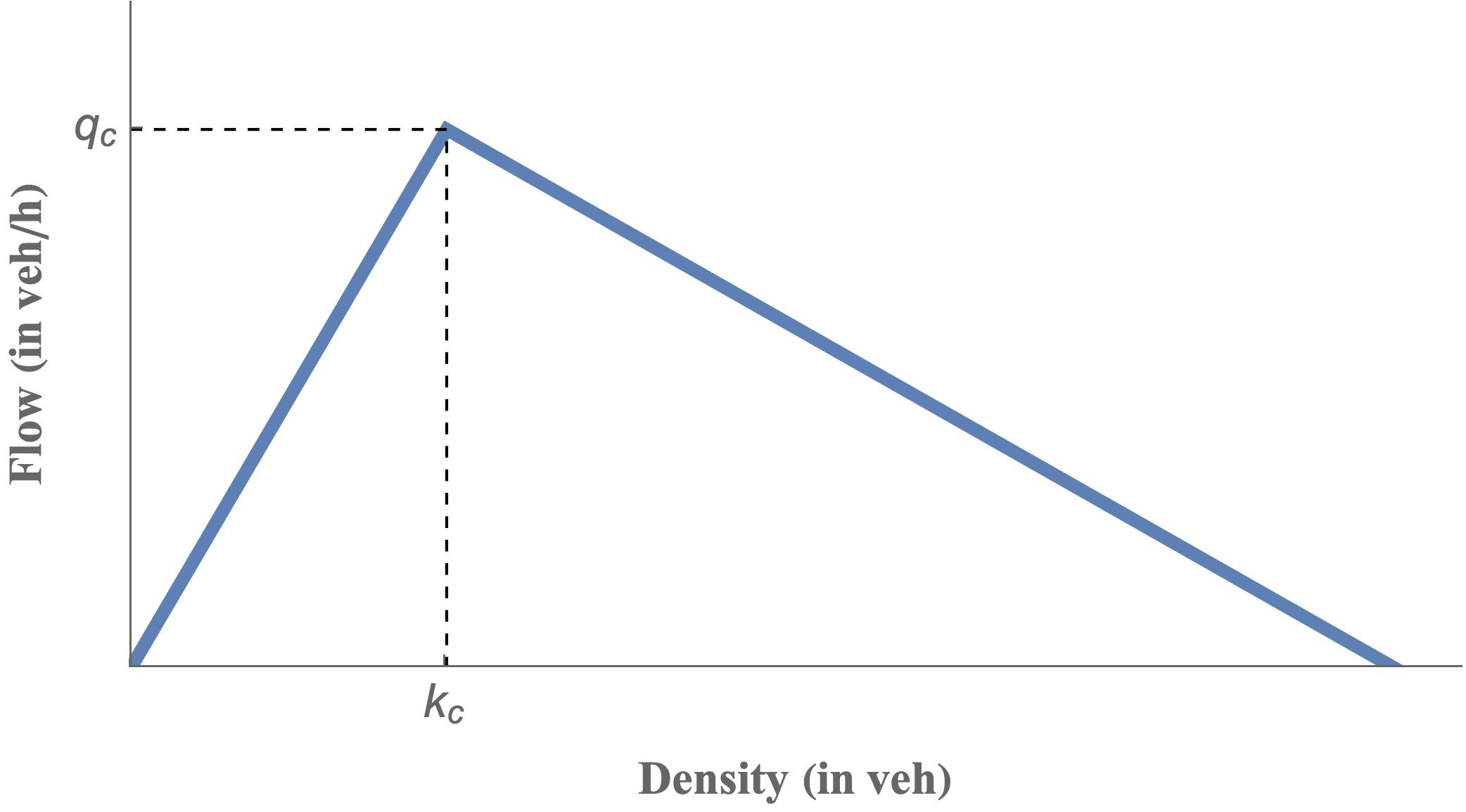}
    \caption{Example of triangular fundamental diagram.}
    \label{fig:fd-example}
    % \vspace{-14pt}
\end{figure}

Given that travel time is inversely proportional to mean speed, assuming each edge length is 1, and capacity flow $q_c$ is the critical density, i.e., the density at capacity $k_c$ and jam density $k_j$, the following formula can derive the critical density (see also \cite{vitale2024cooperative})
\begin{align} \label{eq:fd}
    q = \begin{cases}
            q_c \dfrac{k}{k_c} & \text{for } 0 \leq k \leq k_c, \\
            q_c(1 - \dfrac{k - k_c}{k_j - k_c}) & \text{for } k_c \leq k \leq k_j,
        \end{cases}
\end{align}
where the travel time is $\dfrac{k}{q}$. From \eqref{eq:fd}, it can be seen that the flow increases for $0 \leq k \leq k_c$ and at the left hand-side of the FD until reaching the critical density, while it starts decreasing after that point for $k_c \leq k \leq k_j$, until reaching to zero (see Figure \ref{fig:fd-example}).

\subsection{Re-routing}
Due to the highly dynamic environment of the urban networks, CAVs may need to re-calculate their route, as the traffic conditions may change. To this end, the RDU is responsible for tracking the corresponding changes in the traffic conditions and deliver updated routes to the CAVs. Specifically, the RDU receiving the density information from the coordinators, monitors in real time if an edge is near the critical point. 

We can estimate the time needed to reach the critical flow, based on the rate of density change, which is denoted as $\dfrac{dk}{dt}$. Let $k_0$ be the density at $t=t_0$. Assuming that the density changes at a constant rate, we have

\begin{equation}
  r = \dfrac{dk}{dt},
\end{equation}
where $r$ is the rate of density change measured in veh/km per second. To calculate (or estimate) the rate of change of density, $r$, in a traffic flow context, we can use several methods depending on the available data and the specific situation, e.g., learn $r$ through historical data or estimate it using real-time data.

In our framework, we use real-time data estimation from traffic sensors that provide online density measurements. The rate of change is estimated by monitoring the density at given time intervals $\Delta t$. To this end, if $k(t)$ is the density at time $t$, then 

\begin{equation} \label{eq:r_t}
  r(t) = \dfrac{k(t) - k(t+\Delta t)}{\Delta t}.
\end{equation}

Considering that $r(t)$ is approximately constant over a given time interval $[t_0, t_c]$, then the density at time $t$ is given by

\begin{equation} \label{eq:k_t}
  k(t) = k(t_0) + r(t_0) \cdot (t - t_0).
\end{equation}

To calculate the time needed to reach the critical density, $k_c$, we set $k(t_c) = k_c$ in (\ref{eq:k_t}), and by solving for $t_c$, we have
\begin{align}
  k_c &= k(t_0) + r(t_0) \cdot (t_c - t_0) \\
  t_c &= t_0 + \dfrac{k_c - k(t_0)}{r(t_0)}. \label{eq: t_c}
\end{align}

\begin{figure}[t!]
    \centering
    \includegraphics[width = 0.9\linewidth]{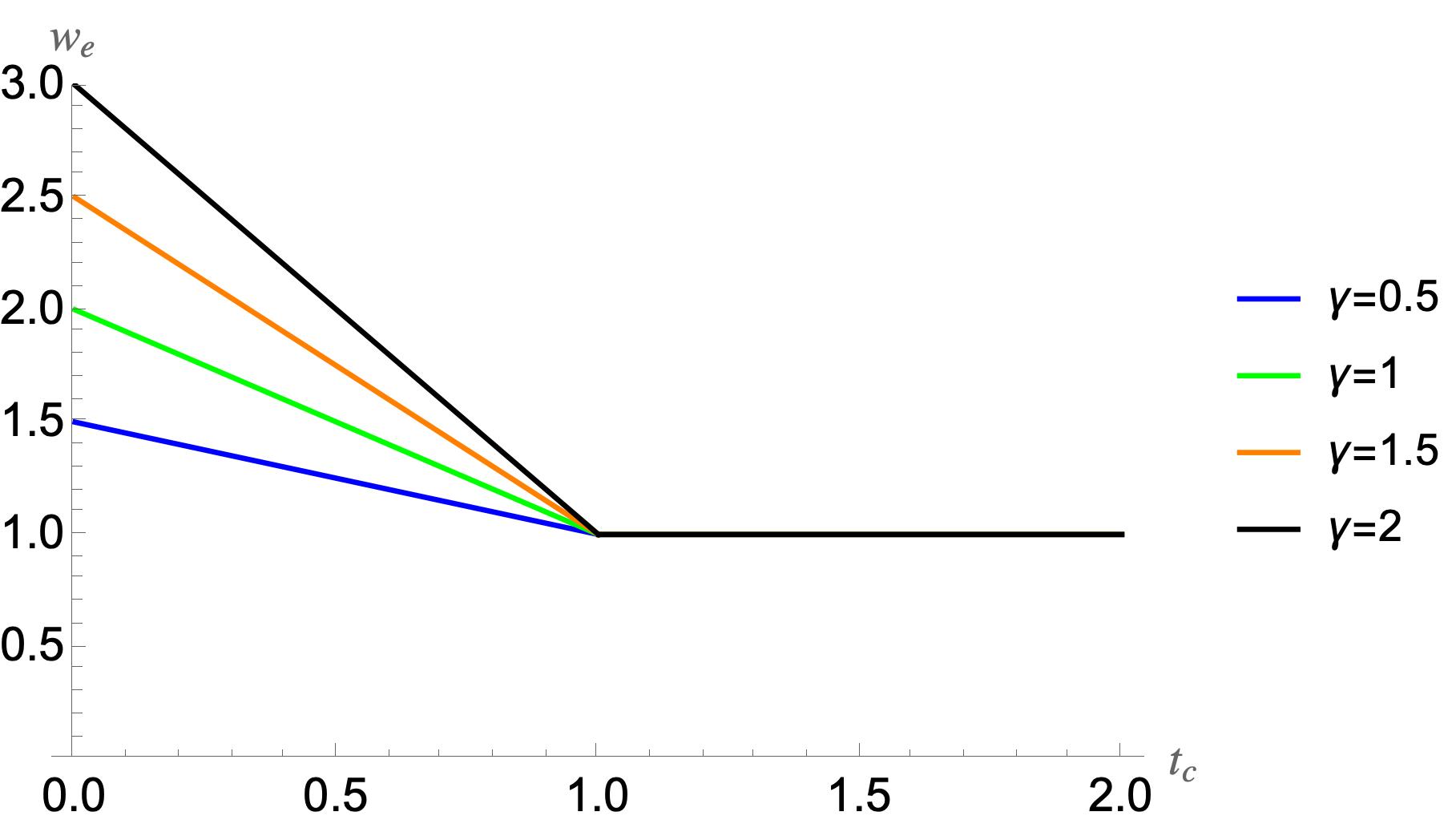}
    \caption{Demonstration of the dynamic weight function for different values of parameter $\gamma$.}
    \label{fig:we}
\end{figure}

Subsequently, we integrate the predicted time to reach the critical density, $t_c$, into the re-routing algorithm by dynamically adjusting the weights of each edge. Thus, the dynamic weight is designed as a function that depends on $t_c$, and features high values when $t_c - t_0$ is less than the threshold, $T_{thres}$, or is equal to a base value, e.g., free-flow travel time otherwise. Based on these requirements, the function of the dynamic weight is
\begin{equation}
  w_e = \begin{cases}
          w_{base} & \text{if } t_c > T_{thres},\\ 
          w_{base} + \gamma \cdot (T_{thres} - t_c) & \text{if } t_c \leq T_{thres},
        \end{cases}
\end{equation}
where $w_{base}$ is the free-flow travel time of each edge, $T_{thres}$ is the time threshold at which the edge is considered to reach its congested state, and $\gamma$ is a scaling parameter. The value of the weight $w_e$ increases as $t_c$ decreases, which means that the corresponding edge is less attractive in the routing problem. Figure \ref{fig:we} shows the dynamic weight for different values of $\gamma$ and setting $T_{thres} = 2.0$ seconds, $w_{base} = 1.0$.

\begin{proposition}
  At each edge of a network with density $k(t)$ and critical density $k_c$, there exists a time threshold $T_{thres}$ such that if $t_c-t_0 < T_{thres}$, the edge should be flagged for re-routing to avoid congestion.
\end{proposition}

Proposition 1 enables a proactive instead of reactive approach, meaning that the system can predict that an edge is likely to become congested, based on current density measurements and the rate of change, instead of waiting until the congestion occurs. This measure also allows for better distribution of the traffic flow across the network, as CAVs can be redirected to to routes with lower density before congestion occurs.

A re-routing should be computed periodically, e.g., every $\Delta t$ seconds or when $t_c$ changes substantially, as the traffic density and flow conditions in urban networks are highly dynamic. Regular re-evaluation ensures that the routing decisions consistently capture the current traffic data, allowing the system to adjust and mitigate congestion as conditions evolve. 

The procedure for the re-routing for each CAV is described as follows:
\begin{enumerate}
  \item Compute $t_c$ for all edges using \eqref{eq: t_c}.
  \item Calculate the weights $w_e$ and apply the chosen weight to each edge.
  \item Update the routing graph by incorporating the new weights into the graph.
  \item Run the routing algorithm to find the optimal paths for CAVs based on the updated weights.
\end{enumerate}

The proposed re-routing procedure achieves proactive congestion management, as the prediction of $t_c$ allows for anticipating imminent congestion at specific edges. Moreover, the dynamic weighting ensures that real-time traffic conditions are taken into account by the routing problem and that the vehicles are efficiently distributed across the network, reducing the possibility of congestion at critical edges.

\begin{proposition}
  Given density measurements with an error bound $\epsilon$, the prediction error $| \Delta t_c |$ in the time $t_c$ to reach the critical density $k_c$ satisfies the following relationship

  \begin{equation}
    | \Delta t_c | = \dfrac{\epsilon}{|r(t_0)|}.
  \end{equation}

  Thus, $| \Delta t_c |$ is proportional to the measurement error $\epsilon$ and is inversely proportional to the absolute value of the rate of density change $| r(t_0) |$.

  Proof: Let the density measurement at $t_0$ have and error $\epsilon$, such that the measured density $k'(t_0)$ deviates from the actual density $k(t_0)$

  \begin{equation}
    k'(t_0) = k(t_0) + \epsilon.
  \end{equation}

  Using $k'(t_0)$ instead of $k(t_0)$ in (7), the predicted $t'_c$ is

  \begin{equation}
    t'_c = t_0 + \dfrac{k_c - k'(t_0)}{r(t_0)}.
  \end{equation}

  The prediction error $\Delta t_c$ is the difference between $t'_c - t_c$, i.e.,
  \begin{equation}
    \Delta t_c = t'_c - t_c.
  \end{equation}
  Substituting $t'_c$ and $t_c$ in the last equation, we obtain
  \begin{equation}
    \Delta t_c = (t_0 + \dfrac{k_c - k'(t_0)}{r(t_0)}) - (t_0 + \dfrac{k_c - k(t_0)}{r(t_0)}),
  \end{equation}
  which yields
  \begin{equation}
    \Delta t_c = - \dfrac{k'(t_0) - k(t_0)}{r(t_0)}.
  \end{equation}
  By substituting (10) in (14), we have
  \begin{equation}
    \Delta t_c = \dfrac{-\epsilon}{r(t_0)},
  \end{equation}
which implies that the magnitude of the prediction error is 
  \begin{equation}
    |\Delta t_c| = \dfrac{\epsilon}{|r(t_0)|}.
  \end{equation}
\end{proposition}

Proposition 2 describes the relationship between measurement errors and prediction accuracy. This relationship shows that (i) predictions are more reliable when density changes fast; and (ii) the slower the traffic conditions are changing, the more precise measurements are needed. 

\begin{theorem}
  If the rate of change $r(t)$ is bounded within a range $[r_{\min}, r_{\max}]$, the time $t_c$ to reach critical density $k_c$ is bounded by

  \begin{equation}
    t_0 + \dfrac{k_c - k'(t_0)}{r_{max}} \leq t_c \leq t_0 + \dfrac{k_c - k(t_0)}{r_{min}}.
  \end{equation}
\textit{Proof}: We know that if $r(t)$ is constant, the time to critical density is given by (7). If $r(t)$ is bounded such that $r_{min} \leq r(t) \leq r_{max}$, the corresponding bound for $t_c$ are
  \begin{equation}
    t_c \geq t_0 + \dfrac{k_c - k_0}{r_{max}} \quad \text{and} \quad t_c \leq t_0 + \dfrac{k_c - k_0}{r_{min}}.
  \end{equation}
  
  Thus, if the rate of change is bounded within a range, we can predict the time to critical density $t_c$ within a bounded range.
\end{theorem}

\section{Intersection Coordination Problem} \label{section:ICP}
This section outlines, for completeness, the coordination problem at intersections in the network, which was initially presented in \cite{chalaki2021RobustGP}. Specifically, we present the formulation and the solution approach of the energy-time optimal trajectory planning problem for each CAV that aims at crossing through the intersection (for more details, please see \cite{chalaki2021RobustGP}).

\subsection{Deterministic Coordination Problem}

\begin{figure}[t!]
    \centering
    \includegraphics[width = 0.7\linewidth]{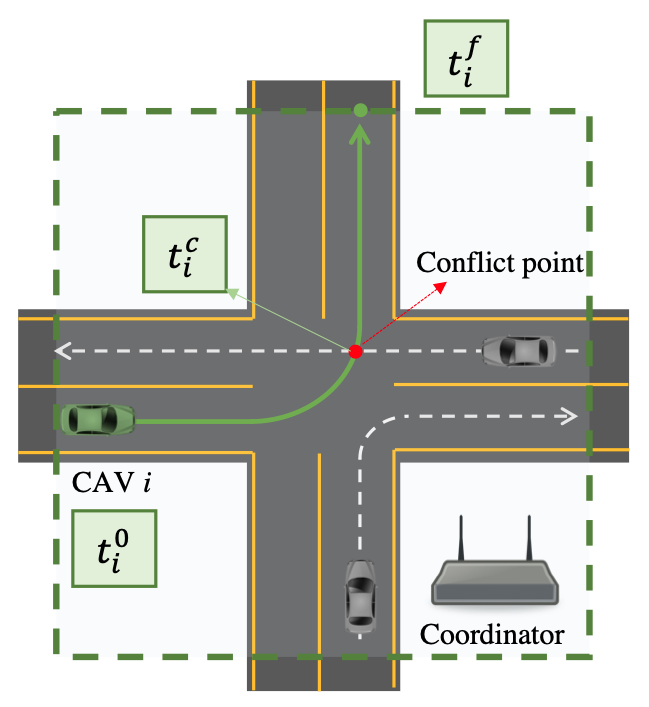}
    \caption{Coordination of connected and automated vehicles (CAVs) at an intersection. CAV $i$ enters at time $t_i^0$, passes the confict point at $t_i^c$, and exits at $t_i^f$.}
    \label{fig:zone}
    \vspace{-14pt}
\end{figure}

Each intersection (see Fig. \ref{fig:zone}) of the network is considered to be signal-free and includes a \textit{coordinator} that acts as a database storing all the information regarding the intersection's geometry and CAVs' trajectories. A \textit{control zone} (illustrated with the dashed line in Fig. \ref{fig:zone}) is considered for each intersection, inside of which the CAVs can communicate with the coordinator.

The dynamics of each CAV $i \in \mathcal{N}(t)$ are described as 
\begin{align}
    \dot{p}_i(t) = v_i(t) \label{state_p}, \\
    \dot{v}_i(t) = u_i(t) \label{state_v},
\end{align}
where $p_i(t), v_i(t)$ are the state variables, denoting the position and the speed of each CAV at time $t$, respectively, while $u_i(t)$ is the control input representing the acceleration. Note that states and acceleration are bounded within feasible regions based on the CAVs' specifications 
\begin{align}
    u_{i,\min} & \leq u_i (t) \leq u_{i,\max}, \\
    0 < v_{\min} & \leq v_i (t) \leq v_{\max},
\end{align}
where $u_{i,\min}$, $u_{i,\max}$ are the minimum and maximum control inputs set based on the capabilities of each CAV $i$. The minimum and maximum allowable speeds of each CAV $i$ are denoted as $v_{\min}$, $v_{\max}$ and are considered constant for all CAVs based on the speed limits of each edge $e \in \mathcal{E}$.

To guarantee rear-end safety between CAV $i$ and the preceding CAV $k \in \mathcal{N}(t) \setminus \{i\}$, the following constraint needs to be satisfied at any time
\begin{equation} \label{eq: rear_col}
    p_k(t) - p_i(t) \geq \delta_i(t) = \gamma + \phi \cdot v_i(t),
\end{equation}
where $\delta_i(t)$ is the safe speed-dependent distance, while $\gamma$ and $\phi \in \mathbb{R}_{>0}$ are the standstill distance and reaction time, respectively.

To avoid lateral collisions between CAV $i$ and CAV $j \in \mathcal{N}(t)\setminus\{i\}$ at conflict point $n$, the following time headway constraint is imposed
\begin{equation} \label{eq: lateral_col}
    | t_i(p_i^n) - t_j(p_j^n) | \geq t_h.
\end{equation}
with $p_i^n$ and $p_j^n$ denoting the distance of the conflict point $n$ from $i$'s and $j$'s paths entries, respectively. Parameter $t_h \in \mathbb{R}_{>0}$ reflects the minimum time headway between any two CAVs crossing conflict point $n$.

\begin{problem} \label{problem_1}
\begin{it}
To determine fuel-efficient trajectories, each CAV $i \in \mathcal{N}(t)$ solves the following optimization problem 
\end{it}
\begin{align*}
    & \min \dfrac{1}{2} \int_{t_i^0}^{t_i^f} u_i(t)^2 dt \\
    & \text{s.t. } \eqref{state_p} - \eqref{eq: lateral_col},
\end{align*}
where $t_i^0$ and $t_i^f$ are the entry and exit times of each CAV $i$ at the control zone, respectively.
\end{problem}
Note that the utilized acceleration cost term $u_i(t)^2$ in the cost criterion was demonstrated in \cite{typaldos2020minimization} to be an excellent proxy for deriving fuel-minimizing vehicle trajectories.

To find the minimum exit time $t^f_i$ , we define the feasible set $\mathcal{T}_i = [\ubar{t}^f_i,\bar{t}^f_i]$, where $\ubar{t}^f_i$ is earliest exit time and $\bar{t}^f_i$ is latest exit time that CAV $i$ can exit the intersection with an unconstrained energy optimal trajectory. This set can be constructed using speed and control input limits (21), (22) and corresponding boundary conditions \cite{chalaki2020experimental}.

\begin{problem}
    To find minimum exit time, each CAV $i \in \mathcal{N}(t)$ solves the following optimization problem
    \begin{align}
        & \min_{t_i^f \in \mathcal{T}_i} t_i^f \nonumber \\
        & \text{s.t.}~ (19)-(24). \nonumber
    \end{align}
\end{problem}

CAV $i$ solves Problem 2 using Algorithm 2. Algorithm 2 returns -1 whenever $t_i^f$ exceeds the maximum allowable exit time with the unconstrained energy optimal trajectory. This implies that there is no unconstrained energy optimal trajectory that satisfies all state, control, and safety constraints.

\begin{algorithm}[H]
\caption{Pseudocode for determining $t_i^f$ of each CAV $i \in \mathcal{N}(t)$}\label{alg:alg1}
\begin{algorithmic}
\STATE 
\STATE \textbf{Input:} $\mathcal{T}_i = [\ubar{t}^f_i,\bar{t}^f_i]$, trajectories of CAVs in $\mathcal{N}(t)$
\STATE \textbf{Output: $t_i^f$}

\STATE 1: $t_i^f \leftarrow \ubar{t}^f_i$
\STATE 2: \textbf{while} $t_i^f < \bar{t}^f_i$ \textbf{do}
\STATE 3: \hspace{0.5cm} $k \leftarrow$ CAV in front of CAV $i$
\STATE 4: \hspace{0.5cm} \textbf{while} $p_k(t) - p_i(t) < \delta_i(t)$ \textbf{do}
\STATE 5: \hspace{1.0cm} $t_i^f \leftarrow t_i^f + dt$
\STATE 6: \hspace{0.5cm} \textbf{end}

\STATE 7: \hspace{0.5cm}$j \leftarrow$ CAV behind of CAV $i$
\STATE 8: \hspace{0.5cm} \textbf{while} $p_i(t) - p_j(t) < \delta_j(t)$ \textbf{do}
\STATE 9: \hspace{1.0cm} $t_i^f \leftarrow t_i^f + dt$
\STATE 10: \hspace{0.5cm} \textbf{end}
\STATE 11: \hspace{0.5cm} $\mathcal{L} \leftarrow$ list of CAVs potential of lateral collision
\STATE 12: \hspace{0.5cm} \textbf{for} $k \in \mathcal{L}$ \textbf{do}
\STATE 13: \hspace{1.0cm} $n \leftarrow$ confict node between $i$ and $k$

\STATE 14: \hspace{1.0cm} \textbf{while} $| t_i(p_i^n) - t_k(p_k^n) | > t_h$ \textbf{do}
\STATE 15: \hspace{1.5cm} $t_i^f \leftarrow t_i^f + dt$
\STATE 16: \hspace{1.0cm} \textbf{end}
\STATE 17: \hspace{0.5cm} \textbf{end}

\STATE 18: \hspace{0.5cm} return $t_i^f$
\STATE 19: \textbf{end}
\STATE 20: return -1
\end{algorithmic}
\label{alg1}
\end{algorithm}

\subsection{Robust Coordination Problem}
For each CAV $i \in \mathcal{N}(t)$, we formulate a robust coordination problem in the presence of uncertainty. We seek to derive the new minimum time $t_i^f$ for CAV $i$ to exit the control zone. This exit time corresponds to a new unconstrained energy-optimal trajectory that satisfies all the state, control, and safety constraints for all realizations of uncertainty. In what follows, let $E_i(\cdot) \subset \mathcal{E}_i, F_i(\cdot) \subset \mathcal{P}_i, G_i(\cdot) \subset \mathcal{V}_i$ denote the bounded confidence intervals of CAV $i$ for random process $e_i(\cdot), f_i(\cdot)$, and $g_i(\cdot)$, respectively.

The rear-end safety constraint (23) is enhanced by incorporating the deviations from nominal position trajectories as follows
\begin{align}
    & \hat{p}_k(t) - \hat{p}_i(t) \geq \hat{\delta}_i(t) = \gamma + \phi \cdot \hat{v}_i(t), \\
    & \forall f_i(t) \in F_i(t), \forall f_k(t) \in F_k(t), \forall g_i(t) \in g_i(t), \nonumber
\end{align}
where $\hat{p}_i(t) = p_i(t) + f_i(t)$, with $f_i(t)$ being the unknown deviation from the nominal position trajectory. The distance between CAV $i \in \mathcal{N}(t)$ and the preceding CAV $k \in \mathcal{N}(t) \setminus\{i\}$ has to be greater than a safe distance $\hat{\delta}_i(t)$ for every realizations of deviations from the nominal trajectories of CAV $i$ and CAV $k$.

Similarty, we enhance the lateral collision constraint \eqref{eq: lateral_col} as follows
\begin{align}
    & \hat{t}_i(p_i^n) - \hat{t}_j(p_j^n) \geq t_h, \\
    & \forall e_i(p_i^n) \in E_i(p_i^n), \forall e_j(p_j^n) \in E_j(p_j^k), \nonumber
\end{align}
to include the CAVs' deviations from their nominal time trajectories. Note that, $\hat{t}_i(p_i) = t_i(p_i) + e_i(p_i)$, with $t_i(p_i)$ being the nominal trajectory solution of Problem 2.

Finally, the deviation of the speed of each CAV $i$ is taken into account by enhancing constraint (22) as follows

\begin{equation}
    v_{\min} \leq \hat{v}_i(t) \leq v_{\max}, \quad \forall g_i(t) \in G_i(t)
\end{equation}
where $\hat{v}_i(t) = v_i(t) + g_i(t)$ and $g_i(t)$ is unknown deviation from nominal speed trajectory.

\begin{problem}
    For each CAV $i \in \mathcal{N}(t)$, the following robust coordination problem is considered

    \begin{align}
        & \min_{t_i^f \in \mathcal{T}_i(t_i^0)} t_i^f \\
        & \text{s.t. } (19),(20), (25)-(27), \nonumber \\
        & \text{and given boundary conditions.} \nonumber
    \end{align}
\end{problem}

\subsection{Limitations}

Algorithm 2 uses an iterative method that incrementally increases $t_i^f$ until either finding a feasible solution or reaching the maximum allowable exit time. When $t_i^f$ exceeds this maximum allowable exit time, the algorithm returns -1, indicating that no unconstrained energy optimal trajectory exists that satisfies all state, control, and safety constraints. To overcome this limitation, for the rare occasions where Algorithm 2 cannot find a feasible solution, we solve the constrained Problem 1 while setting the optimal final time free but weighted with a parameter $w$ (see Problem 4).

\begin{problem} \label{problem_4}
    \begin{it}
    The constrained problem for each CAV $i \in \mathcal{N}(t)$ solves the following optimization problem 
    \end{it}
    \begin{align*}
        & \min \dfrac{1}{2} \int_{t_i^0}^{t_i^f} u_i(t)^2 dt + w\cdot t_i^f \\
        & \text{s.t. } \eqref{state_p} - (24)
    \end{align*}
\end{problem}

Problem \ref{problem_4} considers the constrained problem, where in case of a safety constraint activation \eqref{eq: rear_col}, \eqref{eq: lateral_col}, the vehicle $i \in \mathcal{N}(t)$ needs to be at the conflict point at time $t_i^c = t_i^n + t_h$, i.e., $p_i(t_i^c) = p_i^n$, with $t_i^n$ being the time of the conflict point $n$. 

The solution of this problem can be derived analytically and yields a continuous two-branch piecewise linear function for the optimal control input as follows (for more details, see \cite{papageorgiou2015optimierung}, Chapter 10.5.2, \cite{typaldos2020vehicle})

\begin{equation}
    u_i^*(t) = \begin{cases} 
                c_i^1 t + c_i^2 & t_i^0 \leq t \leq t_i^{c^-}, \\ 
                c_i^5 t + c_i^6 & t_i^{c^+} \leq t \leq t_i^f, 
            \end{cases}
\end{equation}
and, integrating we also derive the optimal speed and position for each $i \in \mathcal{N}(t)$

\begin{equation}
    v_i^*(t) = \begin{cases} 
                \frac{1}{2} c_i^1 t^2 + c_i^2 t + c_i^3 & t_i^0 \leq t \leq t_i^{c^-}, \\ 
                \frac{1}{2} c_i^5 t^2 + c_i^6 t + c_i^7 & t_i^{c^+} \leq t \leq t_i^f, 
            \end{cases}
\end{equation}

\begin{equation}
    x_i^*(t) = \begin{cases} 
        \frac{1}{6} c_i^1 t^3 + \frac{1}{2} c_i^2 t + c_i^3 t + c_i^4 & t_i^0 \leq t \leq t_i^{c^-}, \\ 
        \frac{1}{6} c_i^5 t^2 + \frac{1}{2} c_i^6 t + c_i^7 t + c_i^8 & t_i^{c^+} \leq t \leq t_i^f, 
            \end{cases}
\end{equation}
where $c_i^1,\dots,c_i^8$, are integration constants for each $i \in \mathcal{N}(t)$.

The solution of Problem \ref{problem_4} is more computationally demanding compared to solving Problem 1 in combination with Algorithm 2. However, it always yields a feasible solution.

\section{Simulation Results} \label{sec:simulations}

\begin{figure}[t!]
    \centering
    \includegraphics[width = 0.9\linewidth]{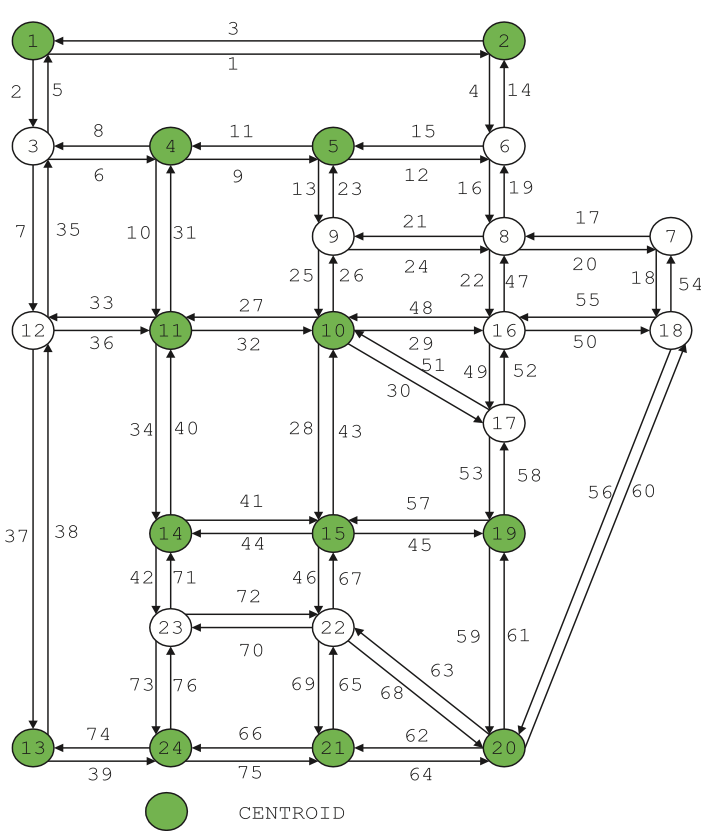}
    \caption{Sioux Falls Network.}
    \label{fig:sioux}
\end{figure}

In this section, simulation results of the proposed approach compared to a baseline approach are presented. The baseline approach considers that CAVs receive a route from RDU only when they enter the network, using the Dijkstra algorithm, and they simply follow their optimal route until they exit the network. For the simulations, the Sioux Falls network was utilized, and the experiments were conducted using Simulation of Urban MObility (SUMO) micro-simulator platform \cite{SUMO2018}. This network contains 24 nodes and 76 links, as illustrated in Figure \ref{fig:sioux}, and is described in Transportation Networks for Research \cite{tnr}.

%%%%%%%%%%%%%%%%%%%%%%%%%%%%%%%%%%%%%%%%%%%%%%%%
\begin{figure*}[!t]
\centering
\subfloat[]{\includegraphics[width=2.5in]{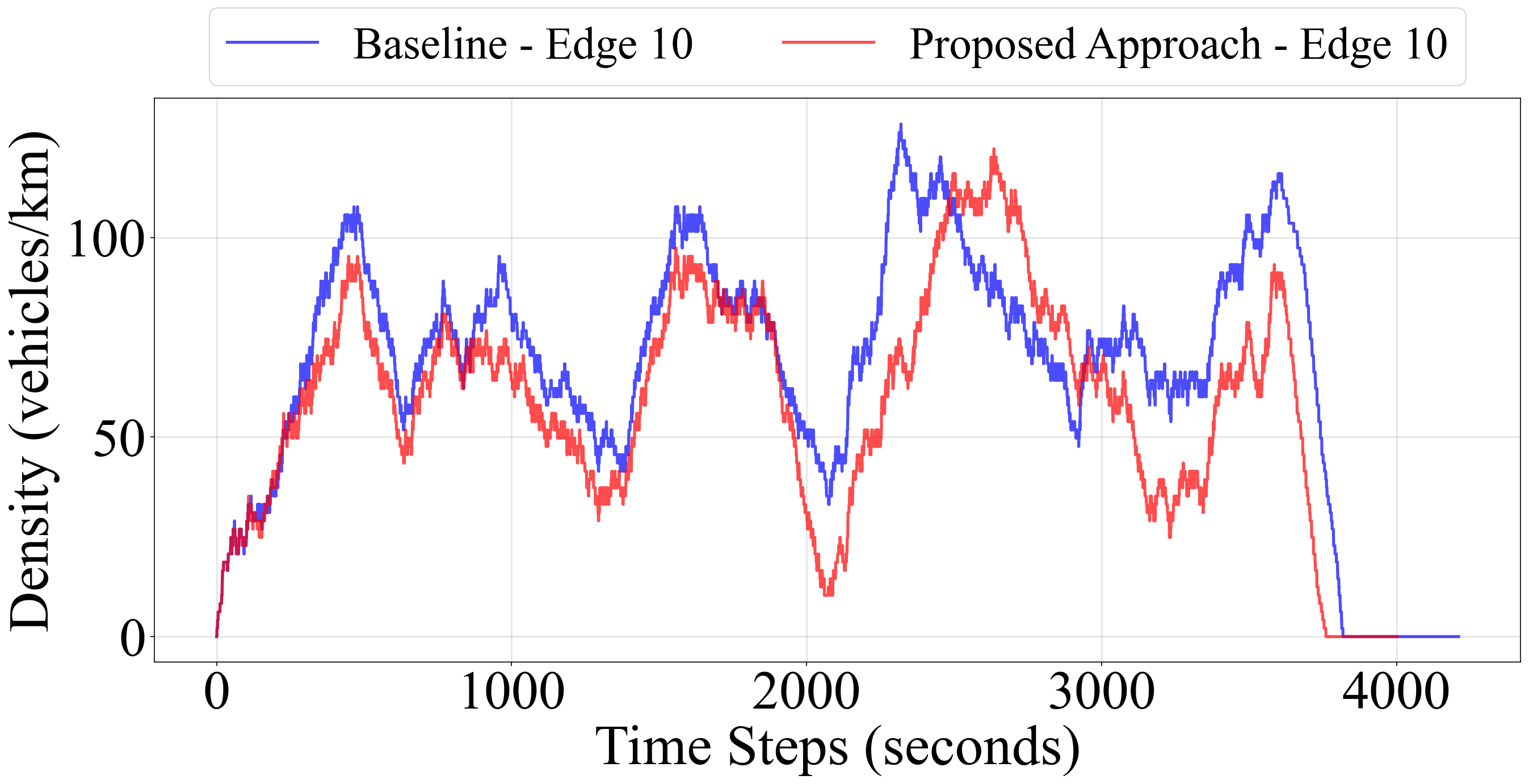}%
\label{fig_first_case}}
\hfil
\subfloat[]{\includegraphics[width=2.5in]{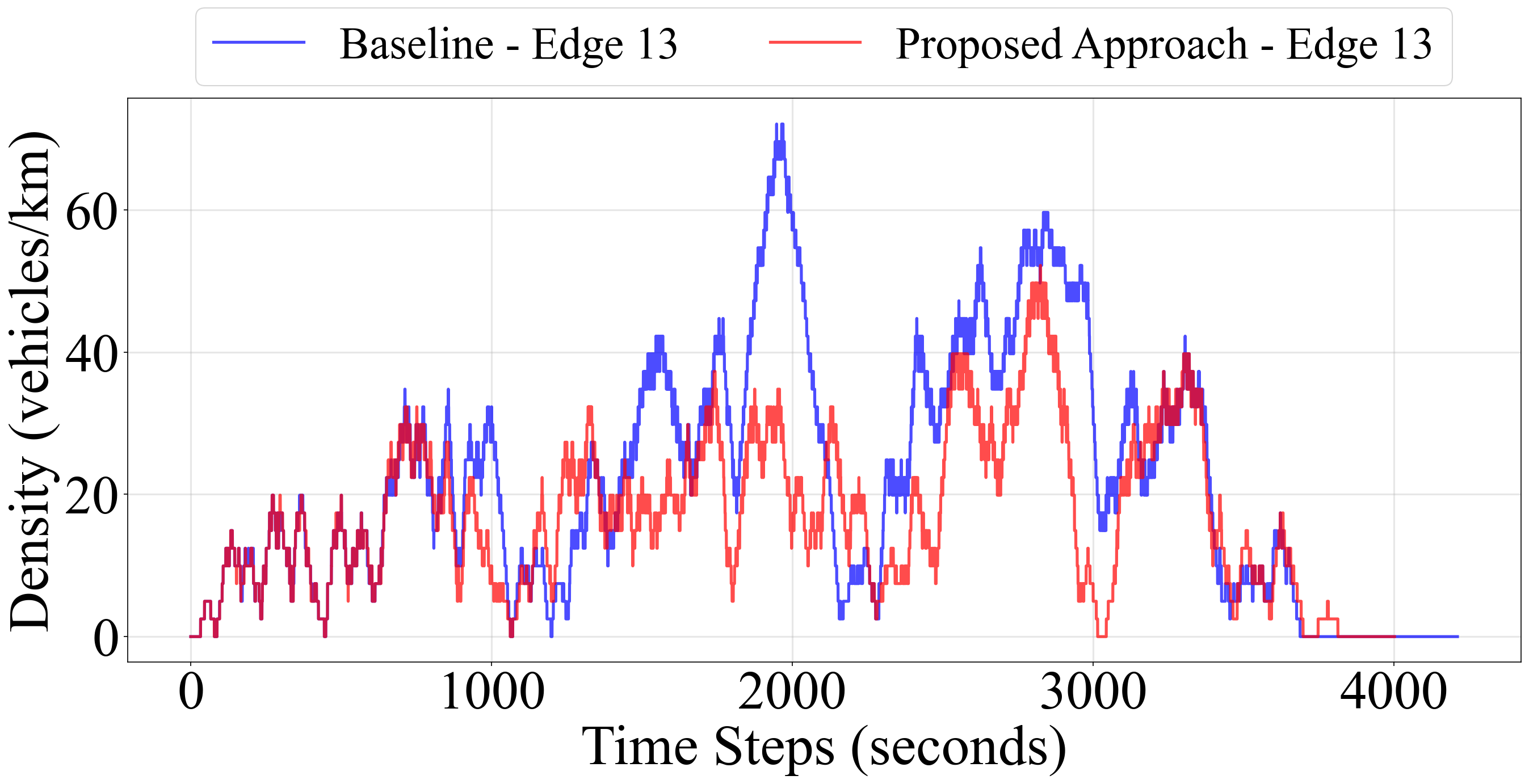}%
\label{fig_second_case}}
\hfil
\subfloat[]{\includegraphics[width=2.5in]{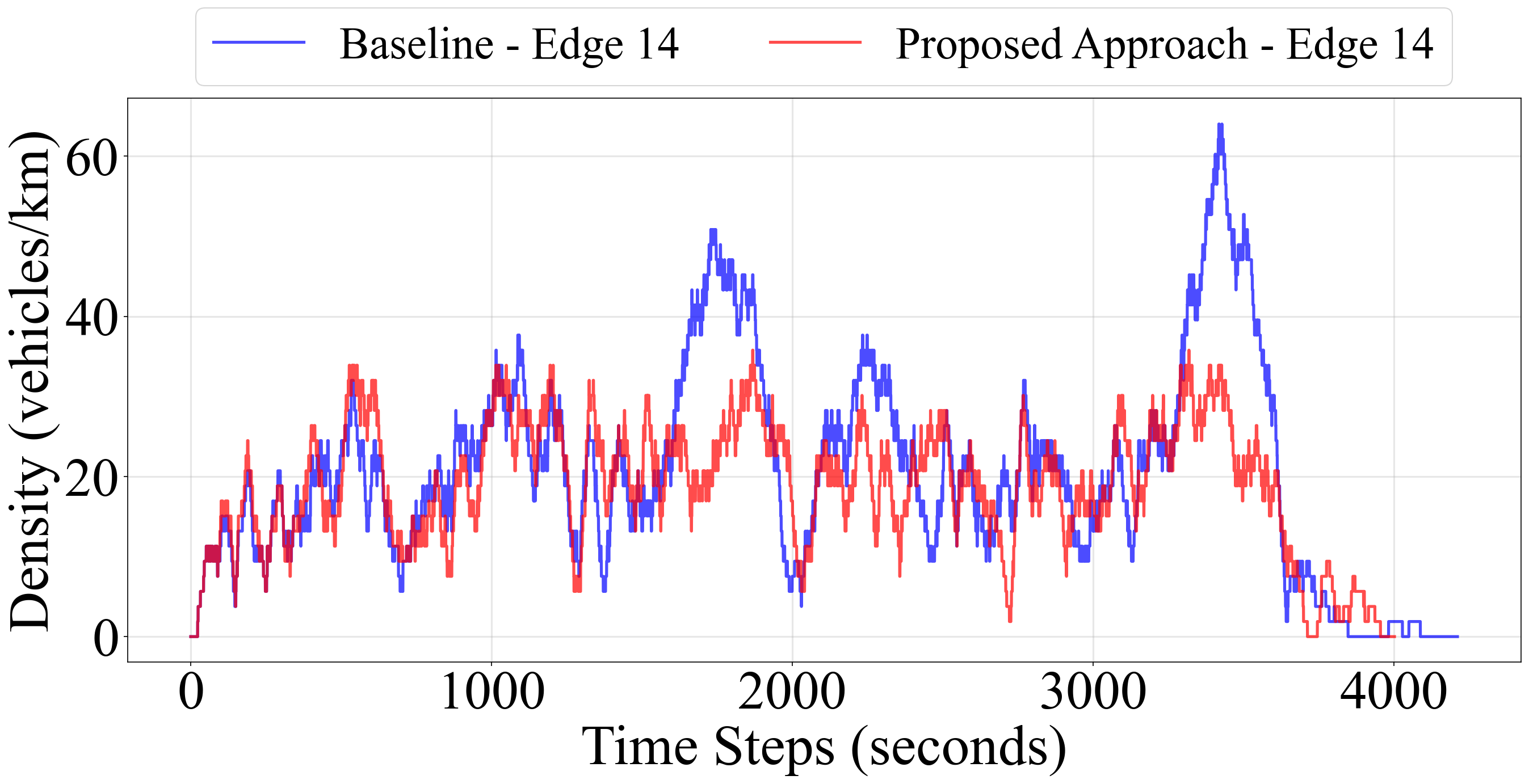}%
\label{fig_second_case}}
\hfil
\subfloat[]{\includegraphics[width=2.5in]{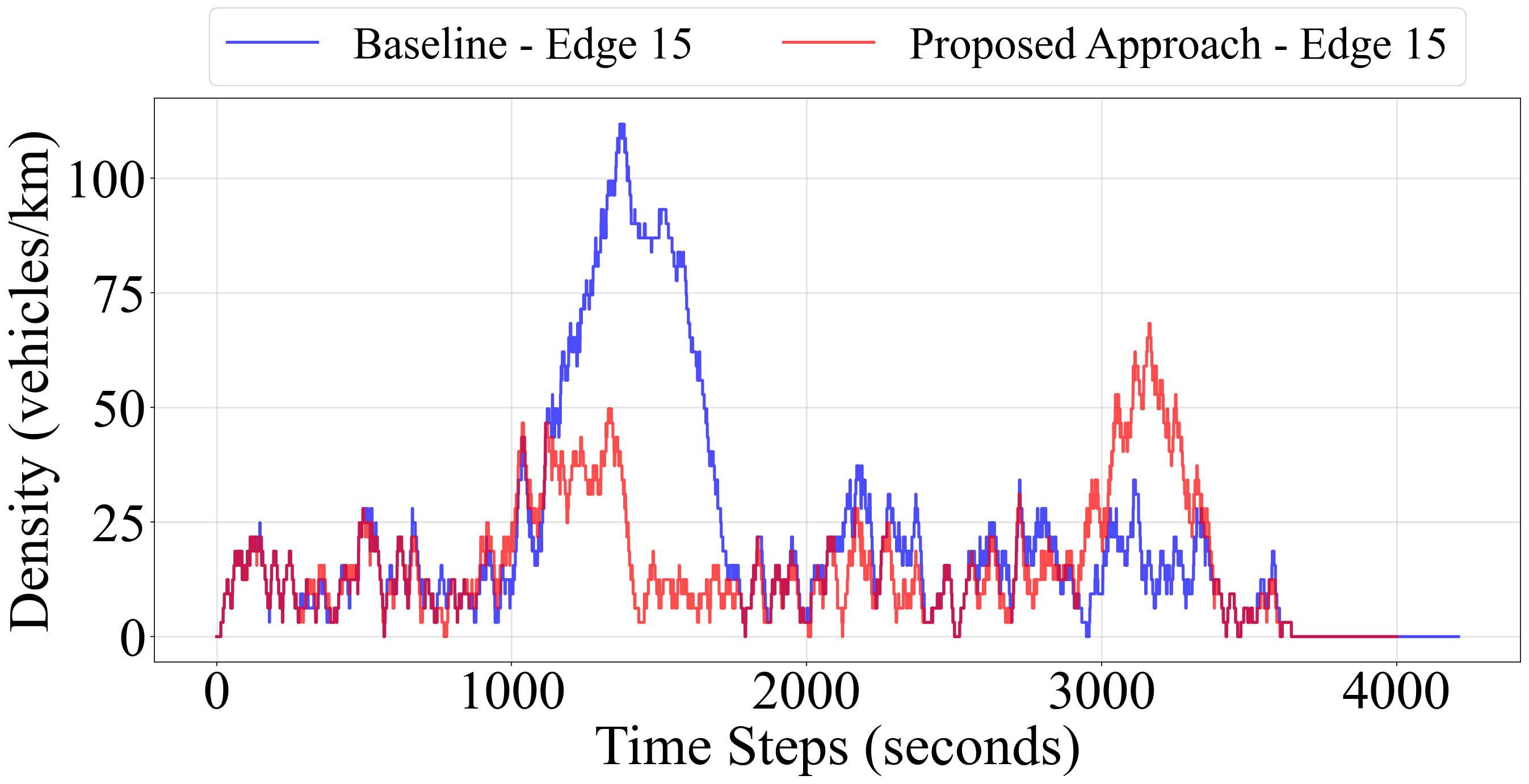}%
\label{fig_second_case}}
\hfil
\subfloat[]{\includegraphics[width=2.5in]{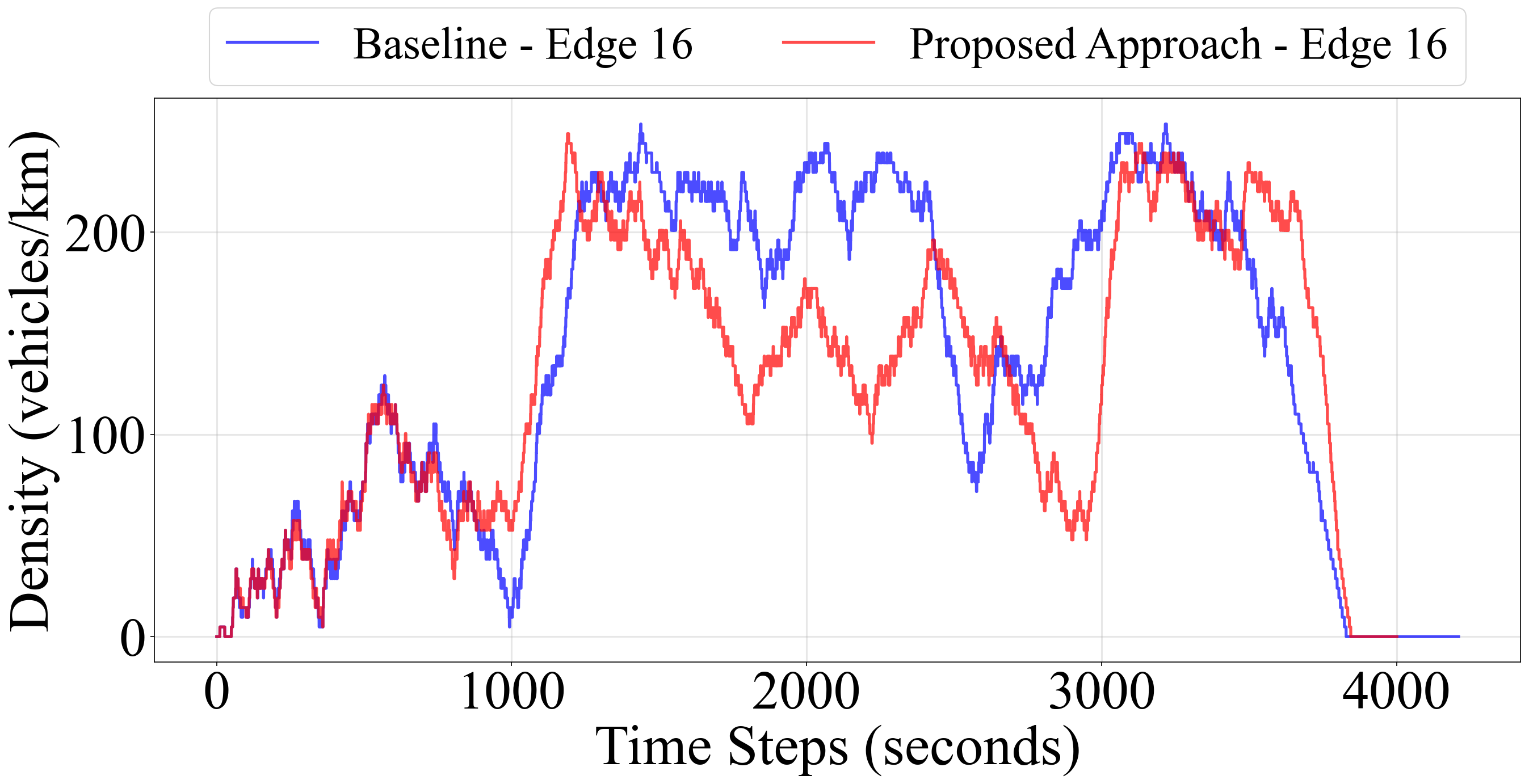}%
\label{fig_second_case}}
\hfil
\subfloat[]{\includegraphics[width=2.5in]{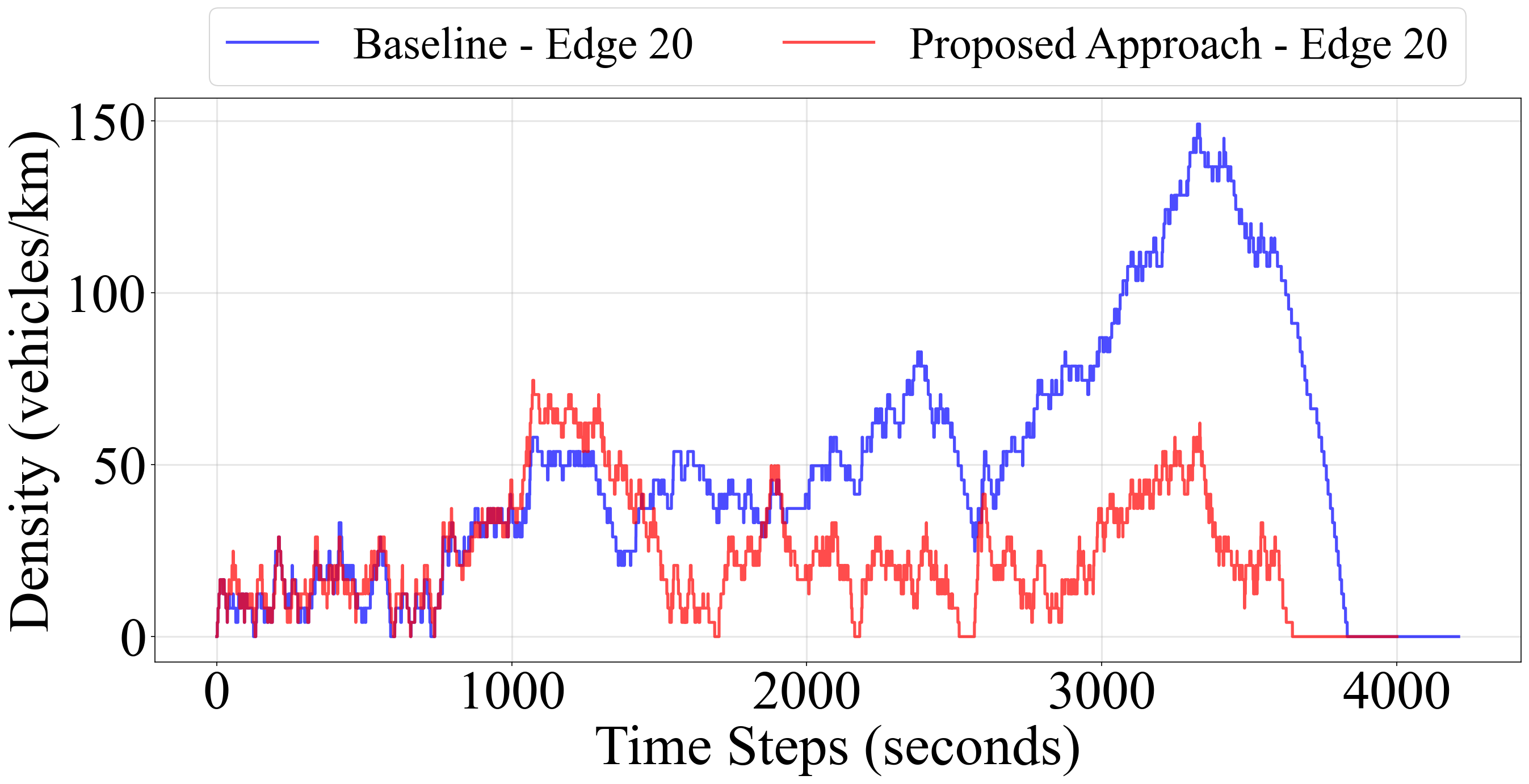}%
\label{fig_second_case}}
\hfil
\subfloat[]{\includegraphics[width=2.5in]{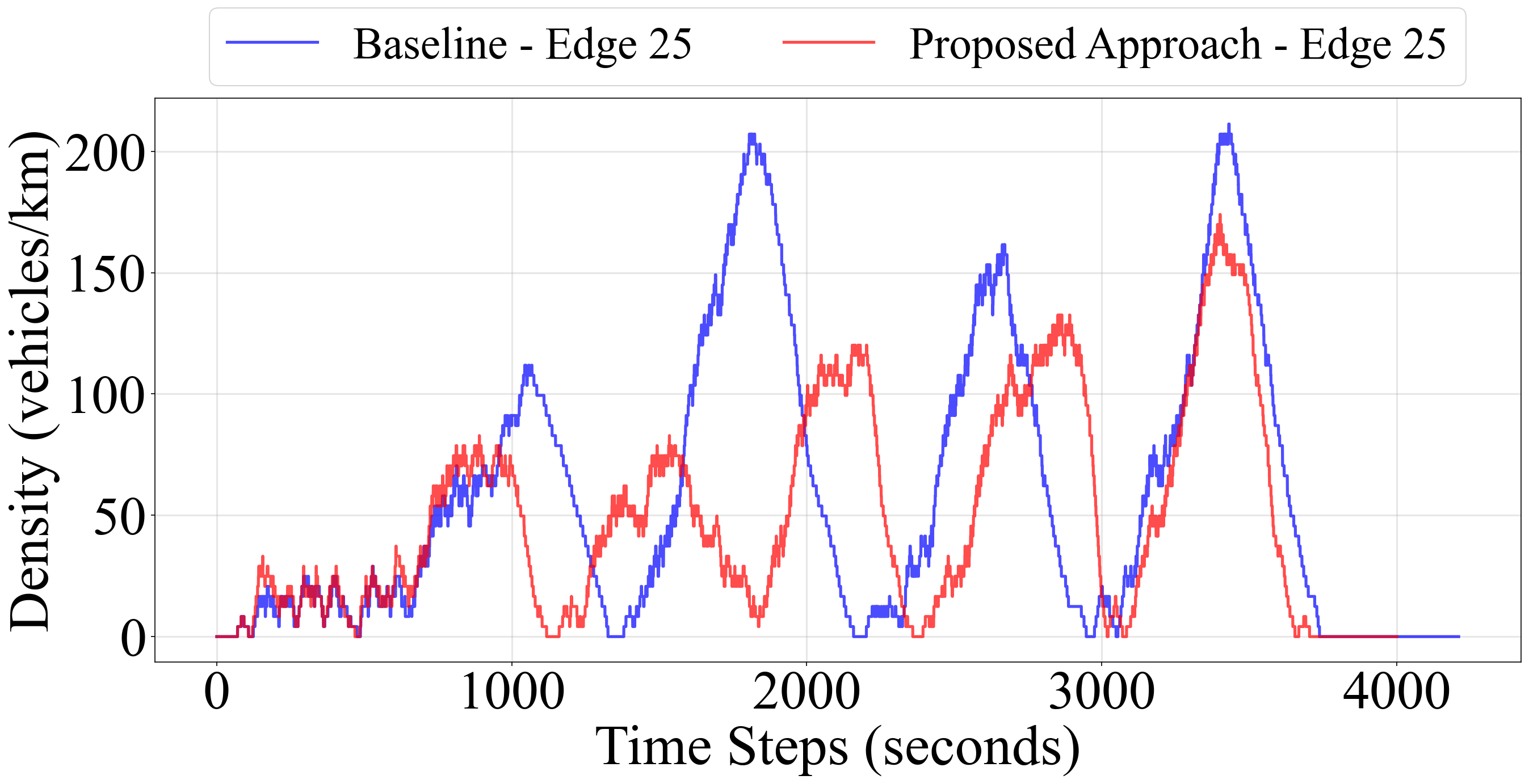}%
\label{fig_second_case}}
\hfil
\subfloat[]{\includegraphics[width=2.5in]{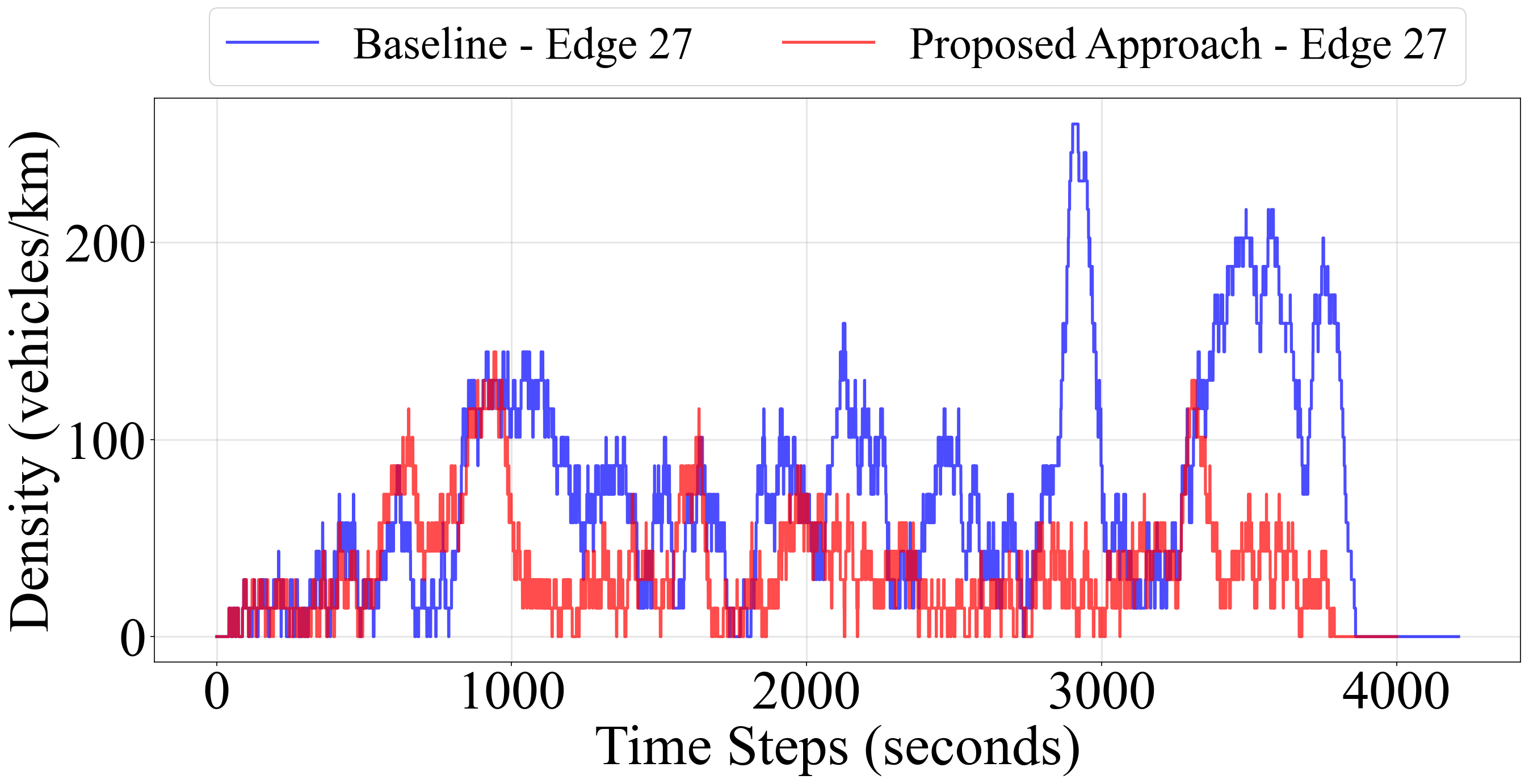}%
\label{fig_second_case}}
\hfil
\subfloat[]{\includegraphics[width=2.5in]{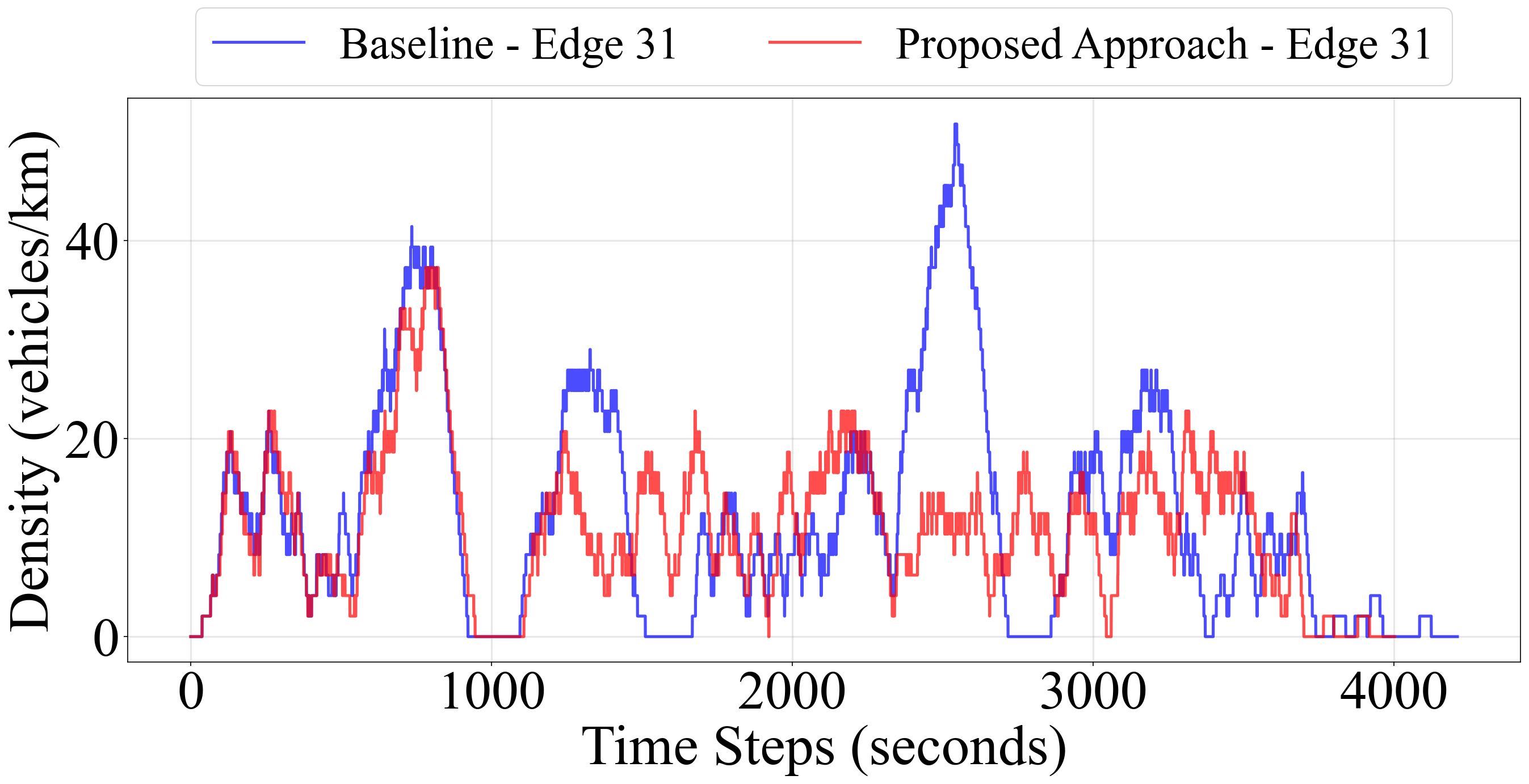}%
\label{fig_second_case}}
\hfil
\subfloat[]{\includegraphics[width=2.5in]{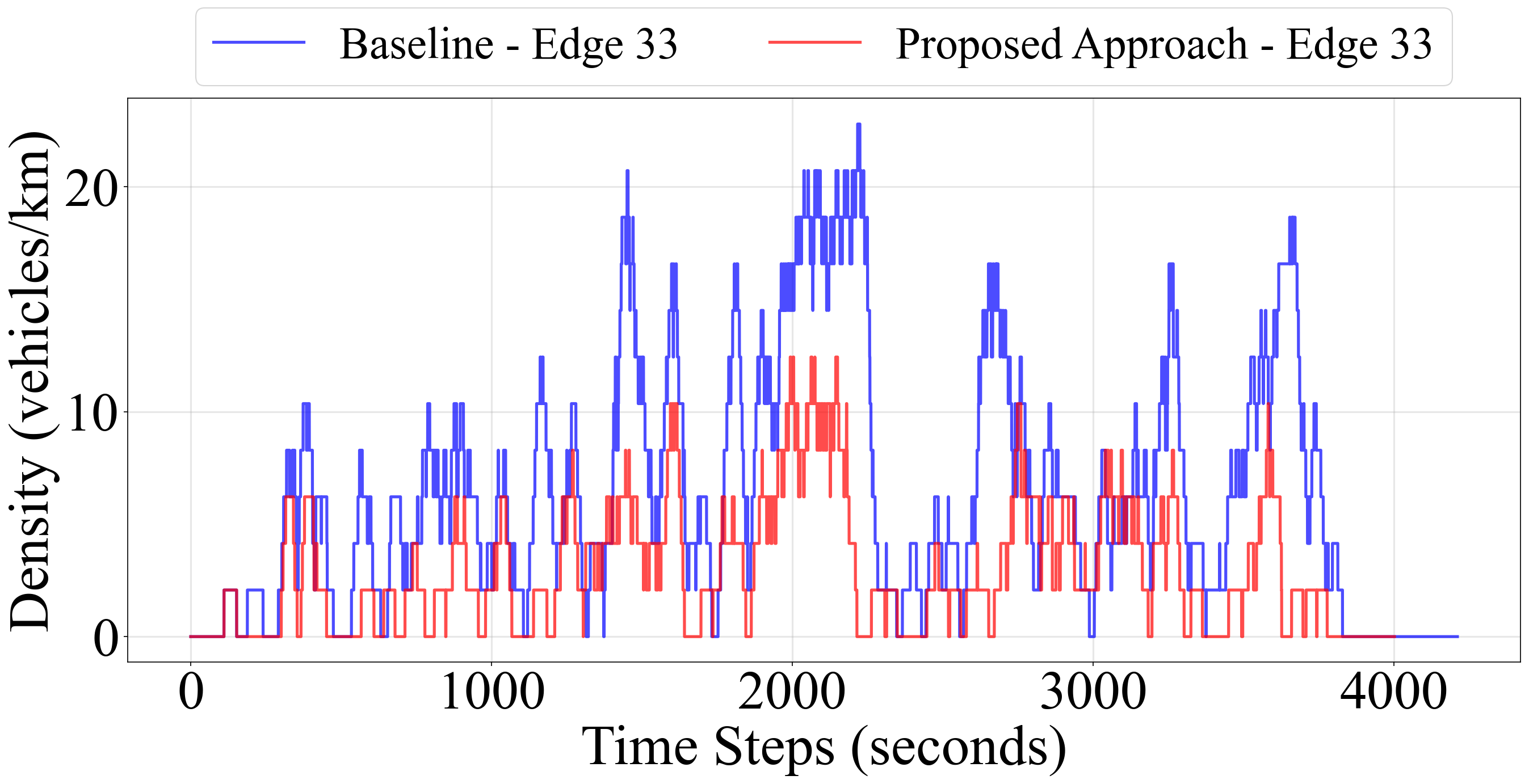}%
\label{fig_second_case}}
\caption{Comparison of density profiles evolution over time between baseline (blue lines) and proposed re-routing approach (red lines).}
\label{fig_sim}
\end{figure*}

Figures 5 and 6 present the density profiles for different edges of the network, where the red lines represent the baseline case and the blue lines show the proposed approach. Specifically, Figure 5 demonstrates the evolution of density over time for different edges of the network. From the figures, it can be seen that the proposed approach consistently achieves lower density values compared to the baseline. This is evident for the subfigures, e.g., (f) and (h), where the proposed approach demonstrates more stable behavior compared to the baseline, which features more fluctuations and higher peaks. This indicates that the proposed approach distributes the traffic more efficiently and consequently manages the congestion points.

The FDs shown in Figure 6 illustrate the relationship between traffic flow and density for different edges in the network, comparing the baseline (blue) and the proposed re-routing approach (red). Similarly to Figure 5, the proposed approach consistently demonstrates superior behavior as it manages to mainly operate in the uncongested area (left side of critical density $k_c$). This is evident from the sublpots, e.g., (i)-(j), where the baseline features extensive values towards the congested area, while the proposed approach manages to mitigate congestion and operate in the uncongested side of the FD. The FDs also reveal that the re-routing strategy effectively prevents capacity drop, as in the majority of the cases, density is under the critical value. This analysis confirms that the proposed re-routing algorithm successfully maintains traffic states in more efficient operating regions by redistributing vehicles before critical density thresholds are reached, thereby avoiding congestion.

The effectiveness of the framework is quantitatively demonstrated through the performance metrics presented in Table I. The proposed approach demonstrates a significant improvement in total travel time (TTT), as it features a significant 12.1\% improvement compared to the baseline. Even more striking is the reduction in total delay (TD), which decreased from 22387.2 to 9899.14 seconds, marking a substantial 55.8\% improvement. This substantial reduction in delay indicates that the proposed re-routing strategy prevents congestion by efficiently redistributing traffic.

Table II shows the environmental impact of the proposed approach and the benefits compared to the baseline. Specifically, it is reported a notable decrease of 10.1\% in fuel consumption. This result is directly correlated with reduced congestion and smoother traffic flow. Additionally, emissions show significant improvements across all pollutant categories, with CO2, CO and NOx been reduced by 10.1\%, 22.45\% and 11.36\% respectively. 

These comprehensive results demonstrate that the proposed hierarchical framework not only improves traffic efficiency through better routing decisions but also yields significant environmental benefits. The substantial reductions in both travel time and emissions suggest that the proactive nature of our approach, which redistributes traffic before congestion occurs, leads to smoother traffic flow throughout the network. The results validate the effectiveness of combining upper-level routing decisions with lower-level trajectory planning, providing a promising solution for future urban traffic management systems.

%%%%%%%%%%%%%%%%%%%%%%%%%%%%%%%%%%%%%%%%%%%%%%%%
\begin{figure*}[!t]
\centering
\subfloat[]{\includegraphics[width=2.5in]{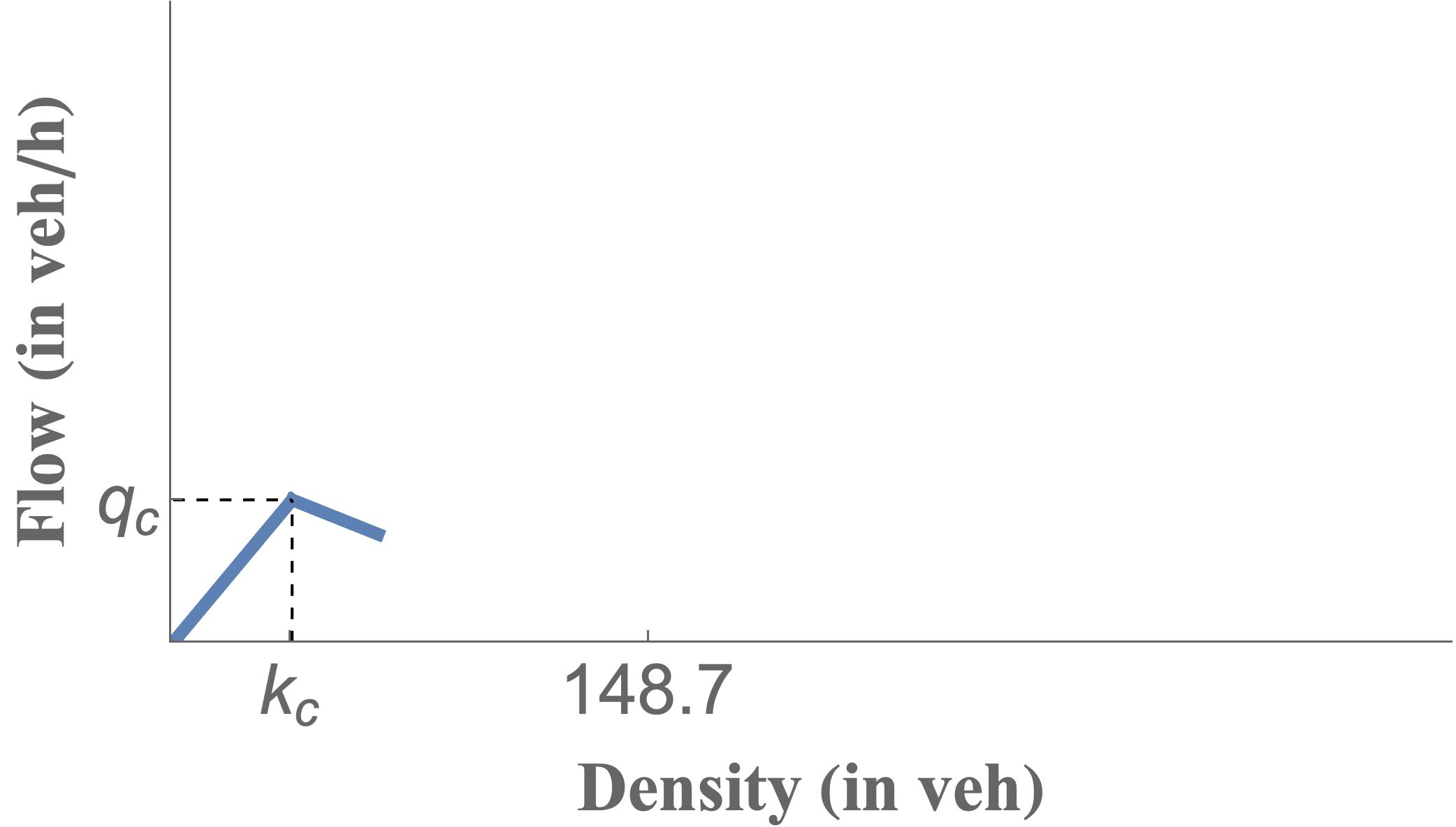}%
\label{fig_first_case}}
\hfil
\subfloat[]{\includegraphics[width=2.5in]{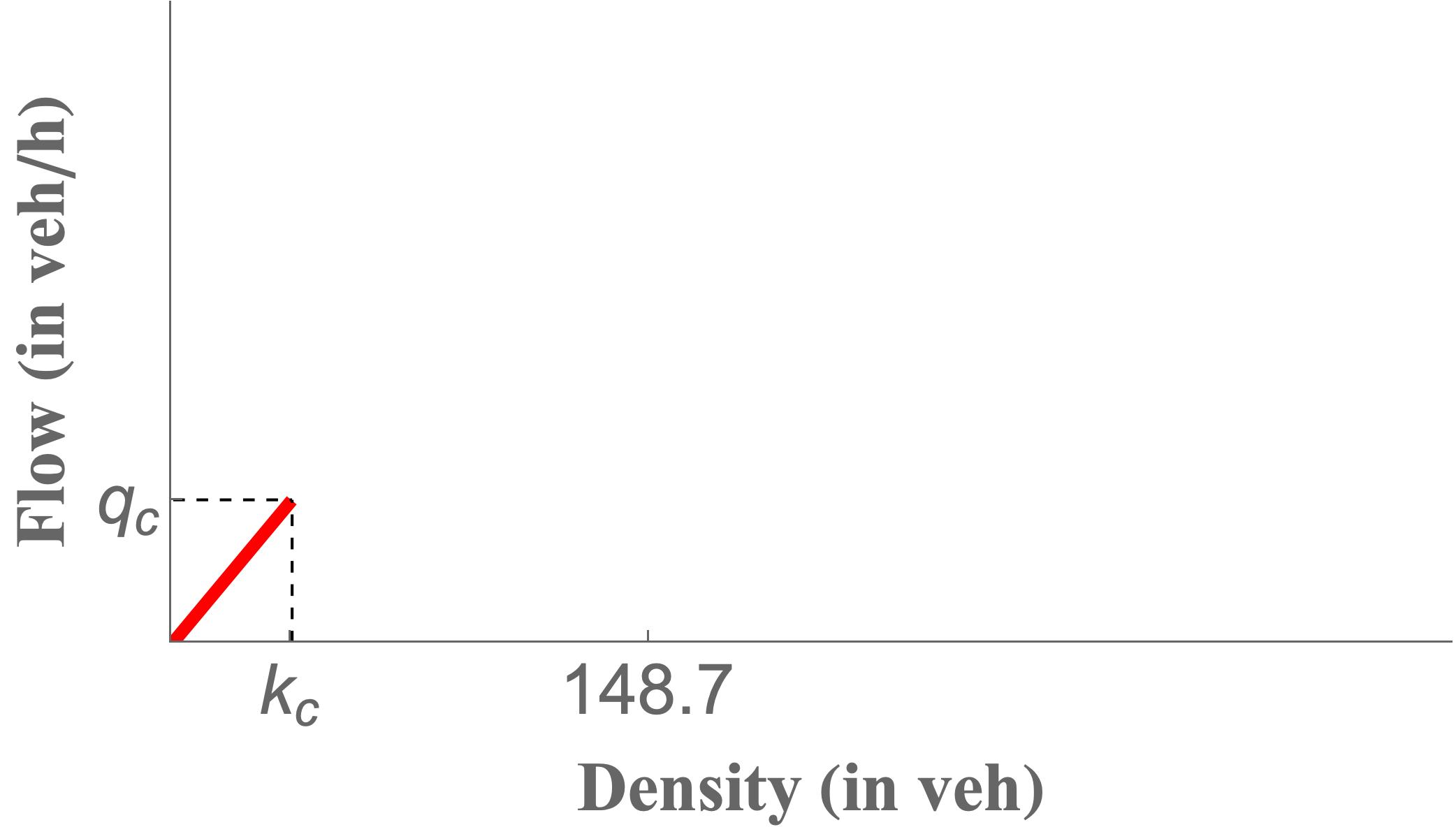}%
\label{fig_second_case}}
\hfil
\subfloat[]{\includegraphics[width=2.5in]{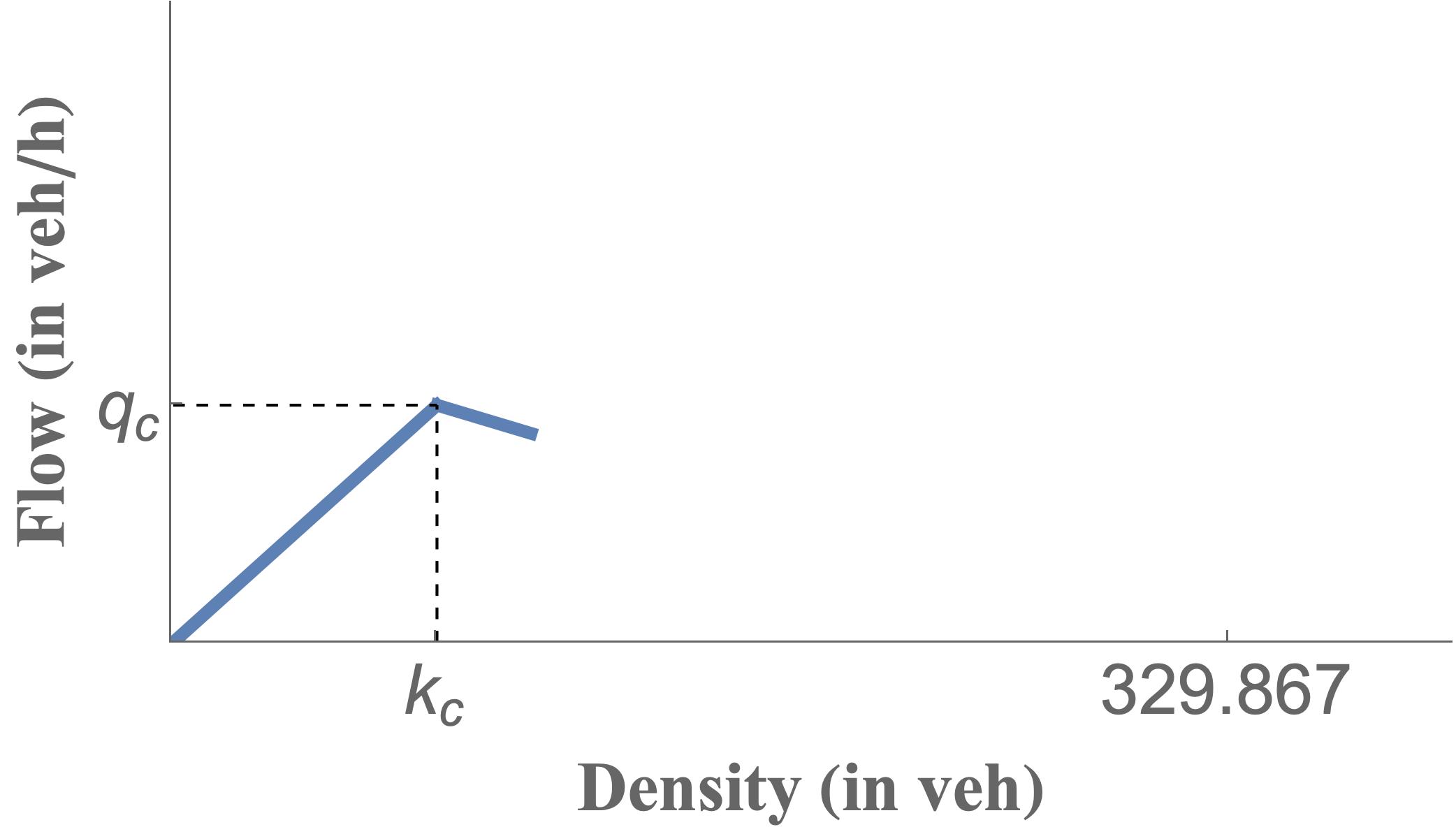}%
\label{fig_first_case}}
\hfil
\subfloat[]{\includegraphics[width=2.5in]{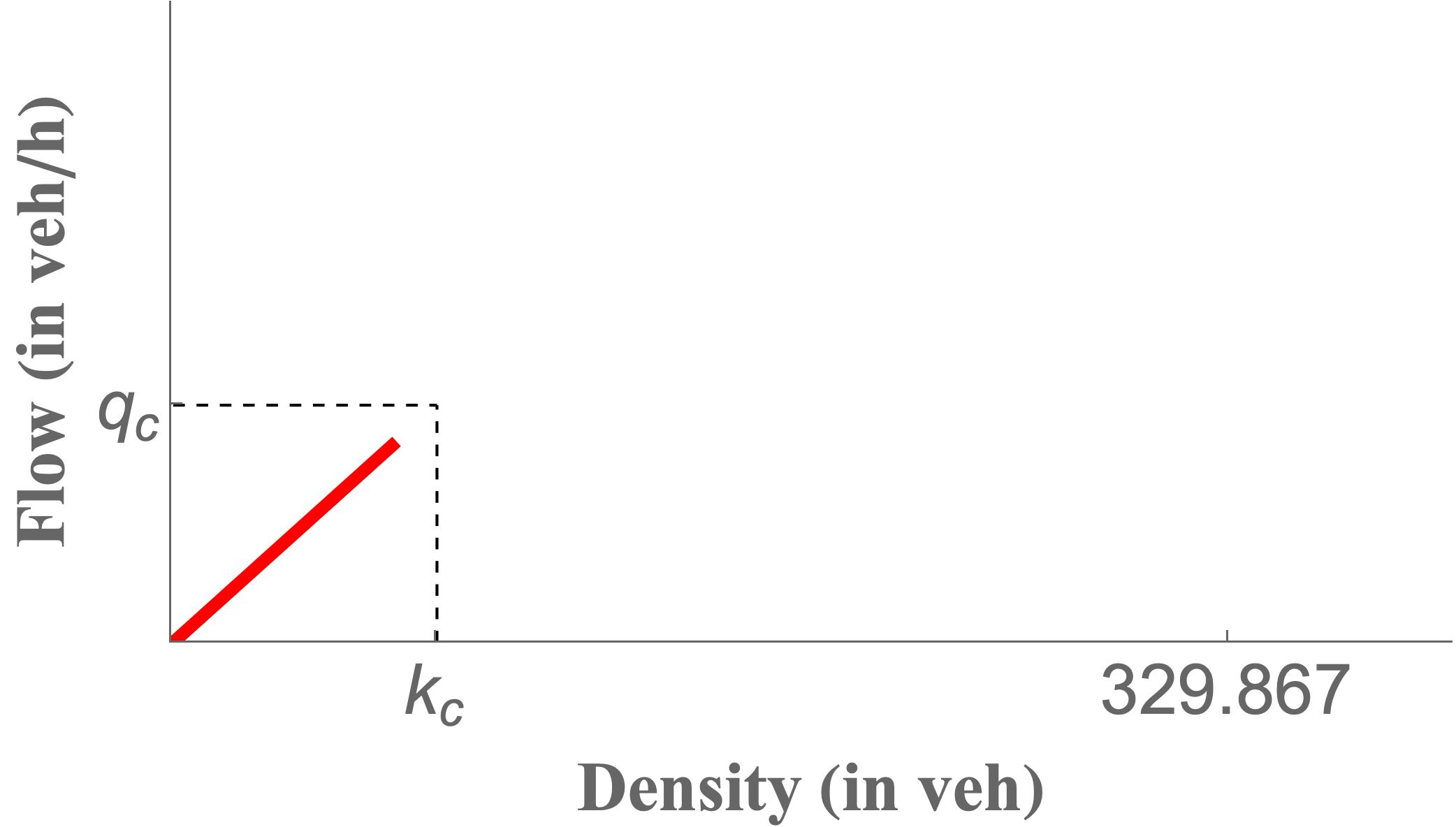}%
\label{fig_second_case}}
\hfil
\subfloat[]{\includegraphics[width=2.5in]{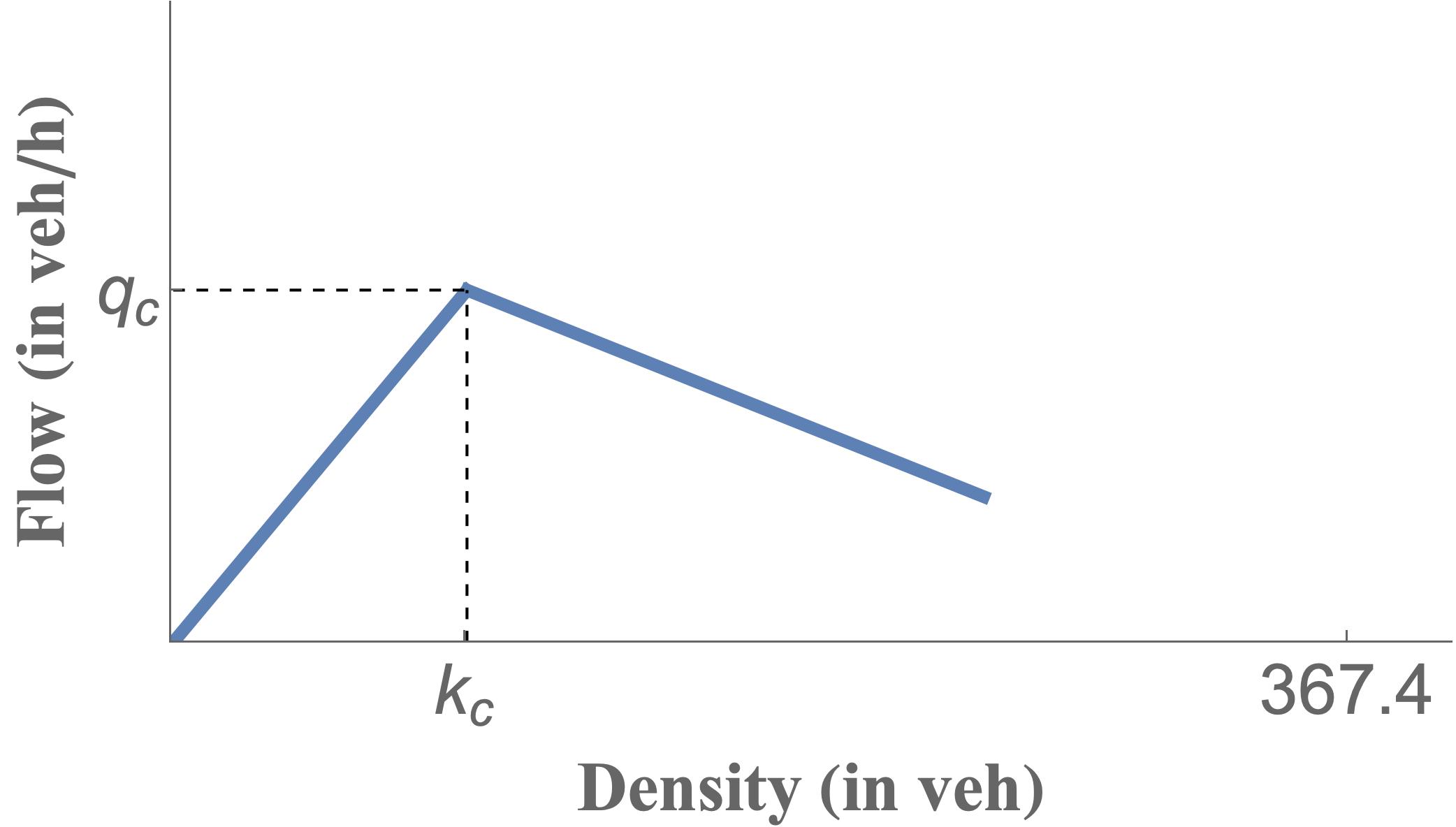}%
\label{fig_first_case}}
\hfil
\subfloat[]{\includegraphics[width=2.5in]{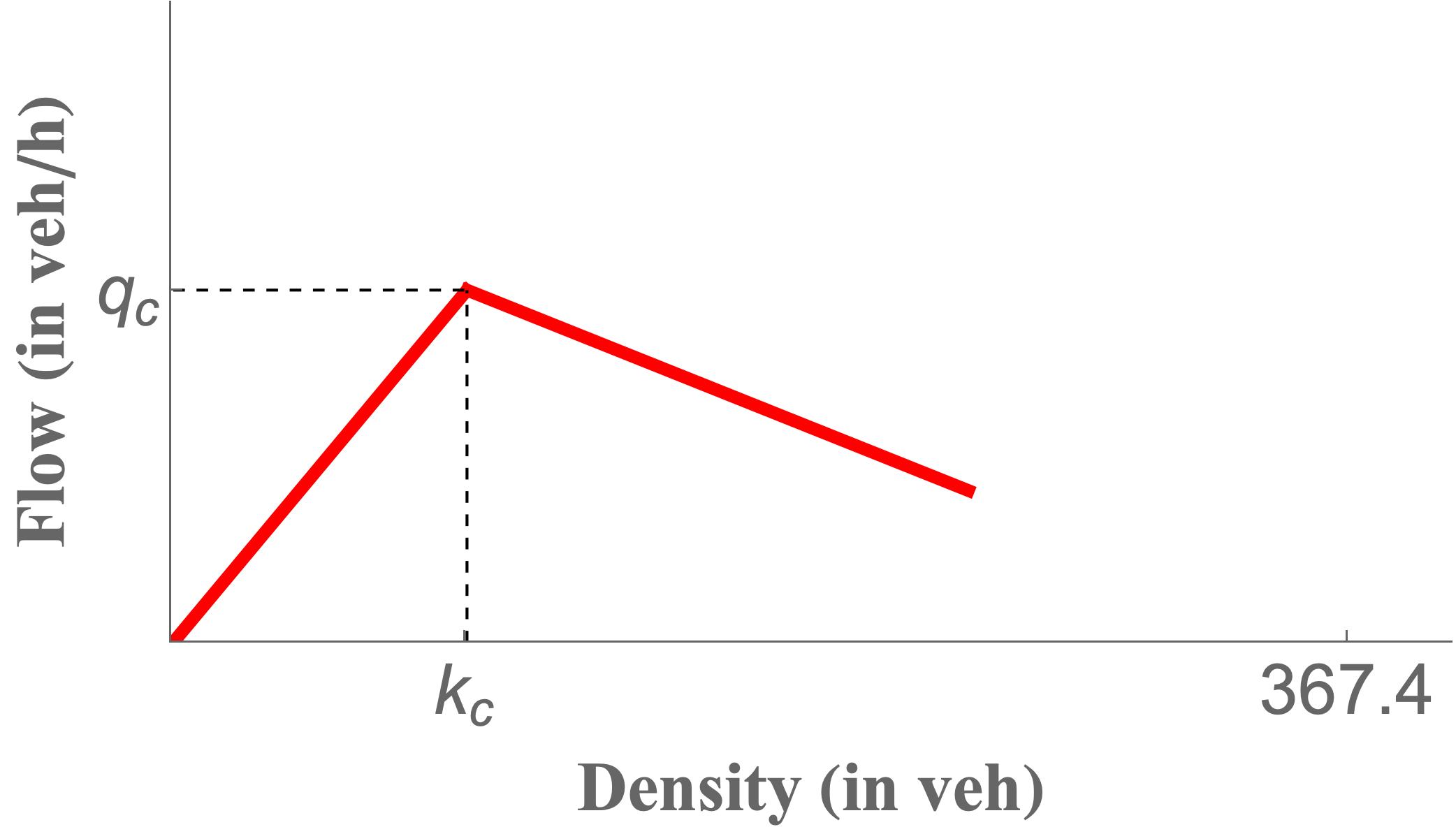}%
\label{fig_second_case}}
\hfil
\subfloat[]{\includegraphics[width=2.5in]{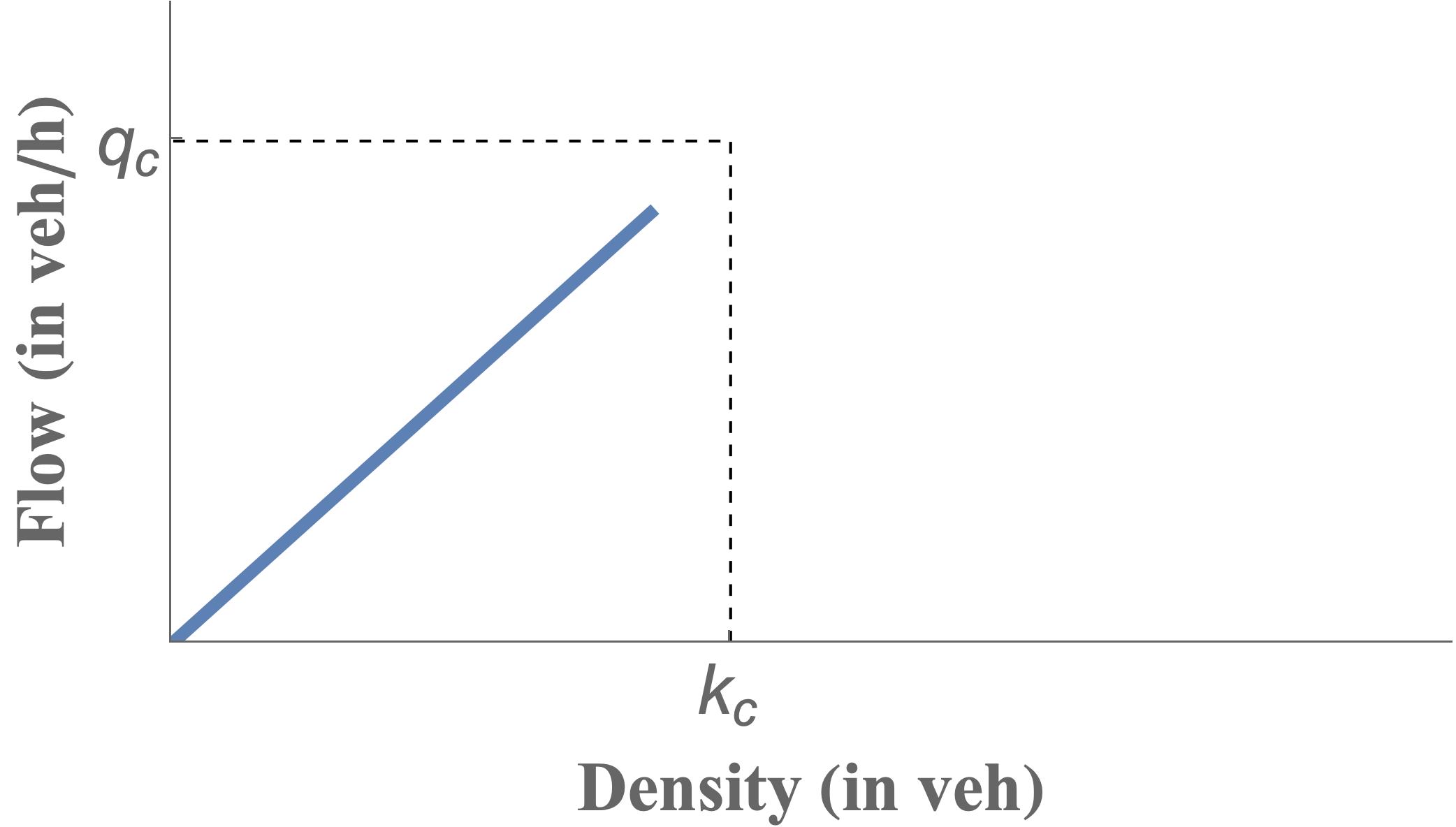}%
\label{fig_first_case}}
\hfil
\subfloat[]{\includegraphics[width=2.5in]{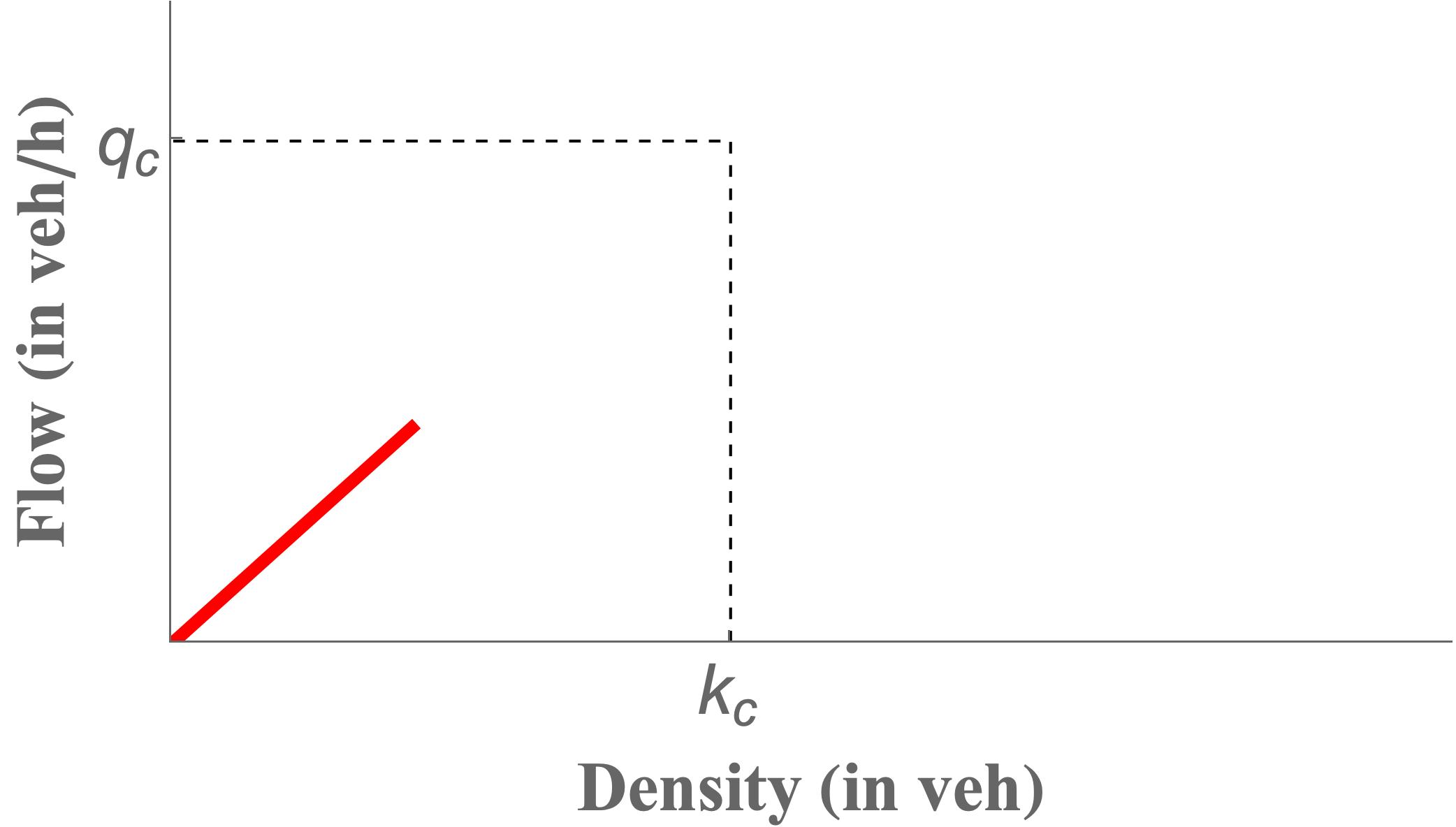}%
\label{fig_second_case}}
\hfil
\subfloat[]{\includegraphics[width=2.5in]{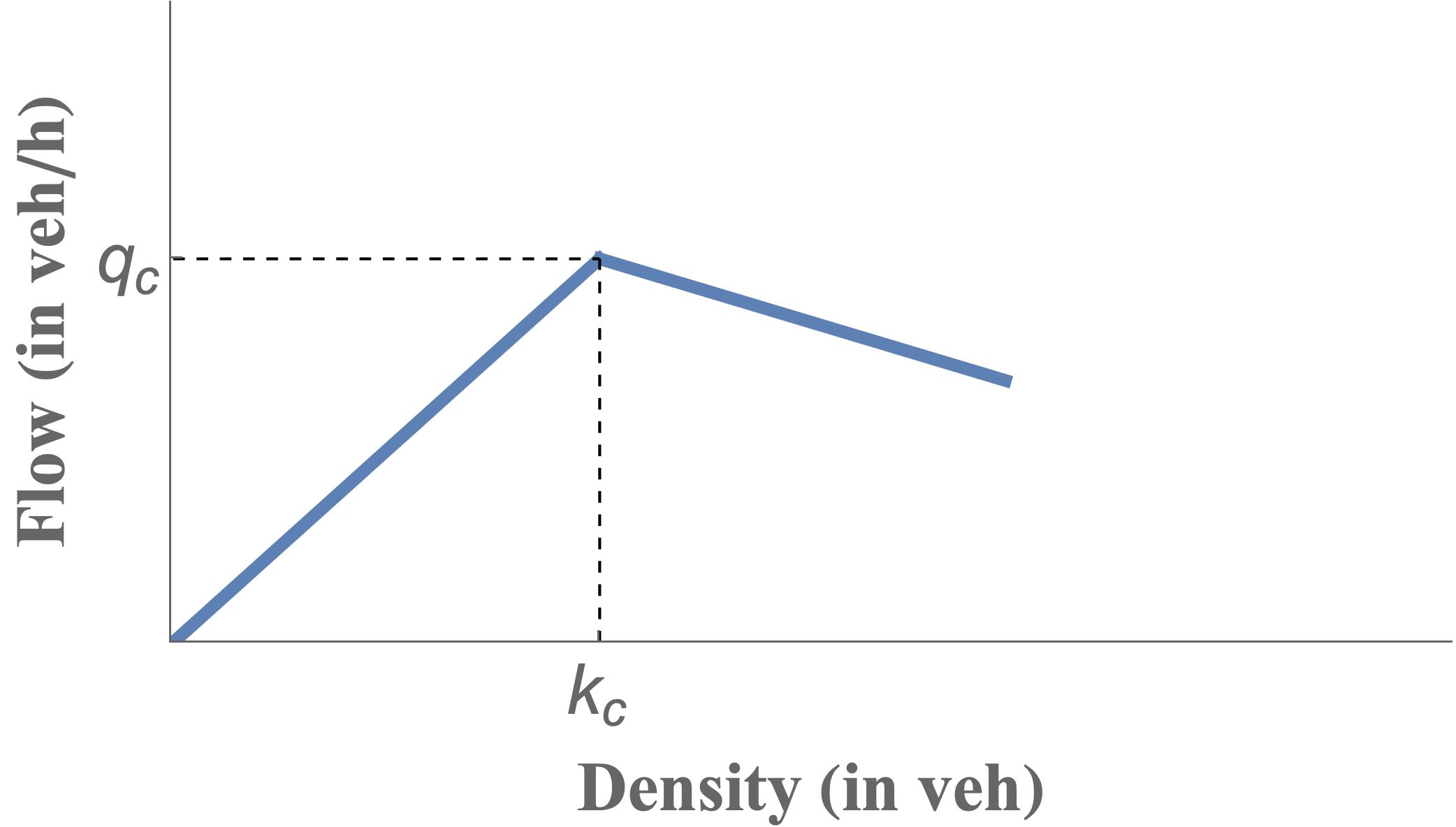}%
\label{fig_first_case}}
\hfil
\subfloat[]{\includegraphics[width=2.5in]{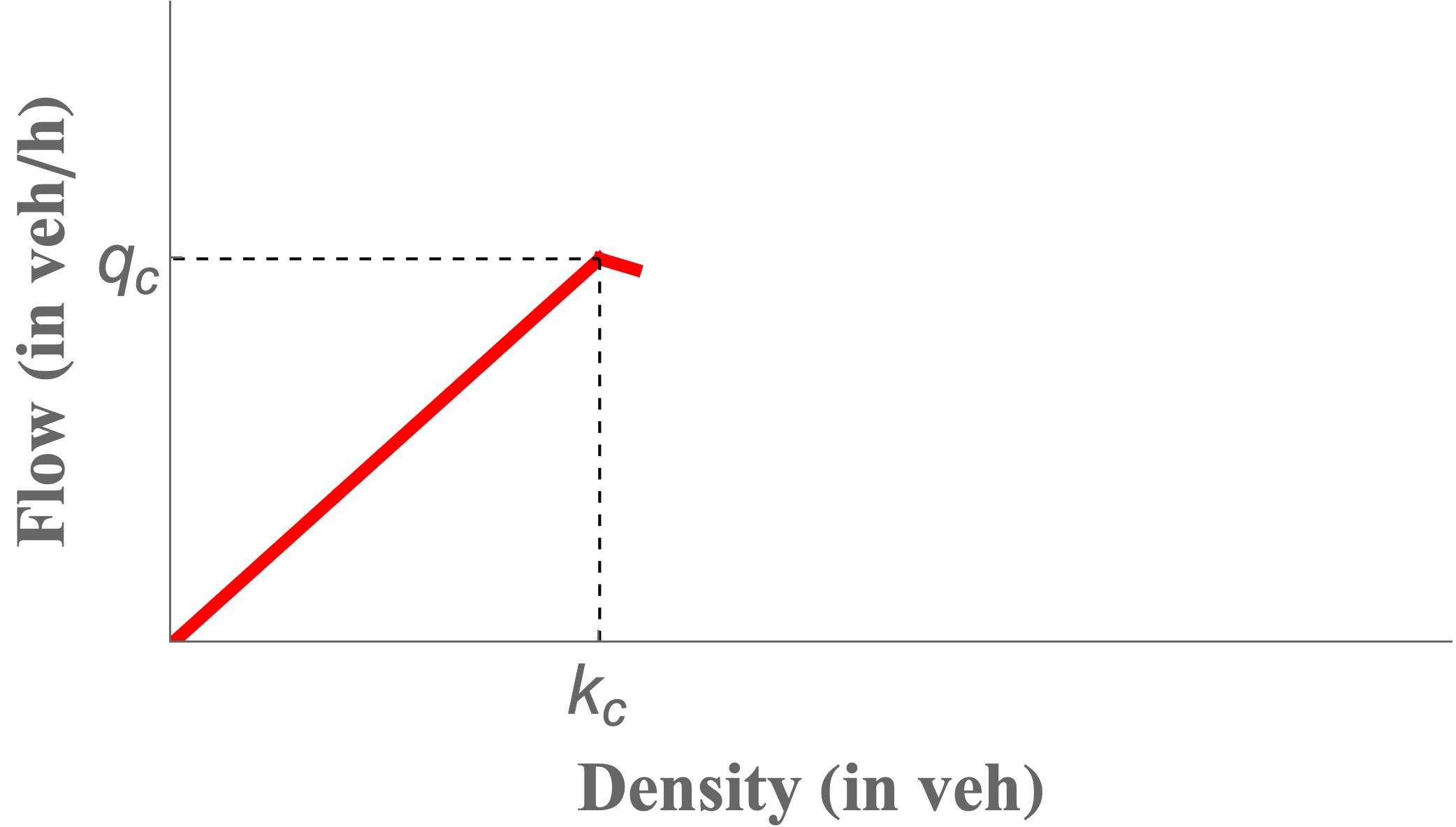}%
\label{fig_second_case}}
\caption{Fundamental diagrams comparing traffic flow-density relationships between baseline (left column, blue) and proposed approach (right column, red).}
\label{fig_sim}
\end{figure*}
%%%%%%%%%%%%%%%%%%%%%%%%%%%%%%%%%%%%%%%%%%%%%%%%
\begin{table}[!t]
\caption{Travel Time and Delay Comparison\label{tab:table1}}
\centering
\begin{tabular}{l c c}
\hline
 & Total Travel Time (TTT) & Total Delay (TD)\\
\hline
Baseline            & 1.47093e+06   & 22387.2  \\
Proposed Approach   & 1.29289e+06   & 9899.14 \\
Improvement \%      & 12.1\%          & 55.8\% \\
\hline
\end{tabular}
\end{table}
%%%%%%%%%%%%%%%%%%%%%%%%%%%%%%%%%%%%%%%%%%%%%%%%    
\begin{table}[!t]
\caption{Emissions and Fuel Comparison\label{tab:table2}}
\centering
\begin{tabular}{l c c c}
\hline
 & Baseline & Proposed Approach & Improvement \% \\
\hline
Fuel Consumption    & 4.5915e+09    & 4.12575e+09   & 10.1\% \\
CO$_2$  & 1.43952e+10   & 1.2935e+10    & 10.1\% \\
CO      & 3.89855e+08   & 3.02333e+08 & 22.45\% \\
NO$_\text{x}$      & 5.89837e+06   & 5.22849e+06 & 11.36\% \\
\hline
\end{tabular}
\end{table}

\section{Concluding Remarks} \label{sec:conclusions}
In this paper, we presented a hierarchical framework that integrates upper-level dynamic re-routing with low-level optimal trajectory planning for CAVs in urban networks. In the upper level, we implemented a dynamic re-routing algorithm that leveraged real-time density information and fundamental diagrams to predict and mitigate potential congestion. By adjusting route weights based on predicted times to reach critical density on each edge, we proactively managed traffic flow. At the lower level, we ensured efficient and safe vehicle crossings at signal-free intersections while maintaining fuel-efficient trajectories through optimal trajectory planning.

Our extensive simulation results on a realistic network confirmed the effectiveness of our approach compared to a baseline method across multiple performance metrics. We successfully reduced total travel time and overall network delay while keeping the density of nearly all edges below the critical threshold. Additionally, we achieved significant environmental benefits, reducing fuel consumption and CO$_2$ emissions by 10.1\%, CO emissions by 22.45\%, and NOx emissions by 11.36\%. These findings validated our framework’s ability to anticipate and prevent congestion while optimizing vehicle trajectories at intersections, ultimately enhancing overall network performance.

Potential directions for future research include investigating the scalability of the framework to larger networks and its performance under mixed traffic conditions at different penetration and compliance rates for human drivers.

\bibliographystyle{IEEEtran}
\bibliography{References, IDS}

% \newpage

 

% \end{IEEEbiography}



% \vspace{-33pt}

\begin{IEEEbiography}[{\includegraphics[width=1in,height=1.25in,clip,keepaspectratio]{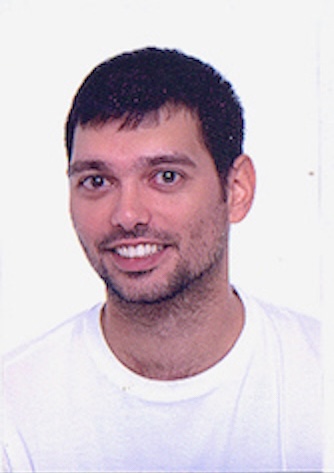}}]{Panagiotis Typaldos} received the B.S. degree in Applied Mathematics from the University of Crete, Heraklion, Greece in 2014 and the M.Sc. degree in Operational Research from the School of Production Engineering and Management, Technical University of Crete (TUC), Chania, Greece, in 2017. In 2022, Dr. Typaldos received his Ph.D. degree at TUC working on the effect of automated vehicles on highway traffic flow and signalized junctions. From May 2024, he is a postdoctoral researcher at the Information and Decision Science Laboratory (IDS) and a Visiting Instructor at Cornell University. His main research interests include optimal control and optimization theory and applications to intelligent transportation systems and trajectory planning for automated ground vehicles.
\end{IEEEbiography}

\begin{IEEEbiography}[{\includegraphics[width=1in,height=1.25in,clip,keepaspectratio]{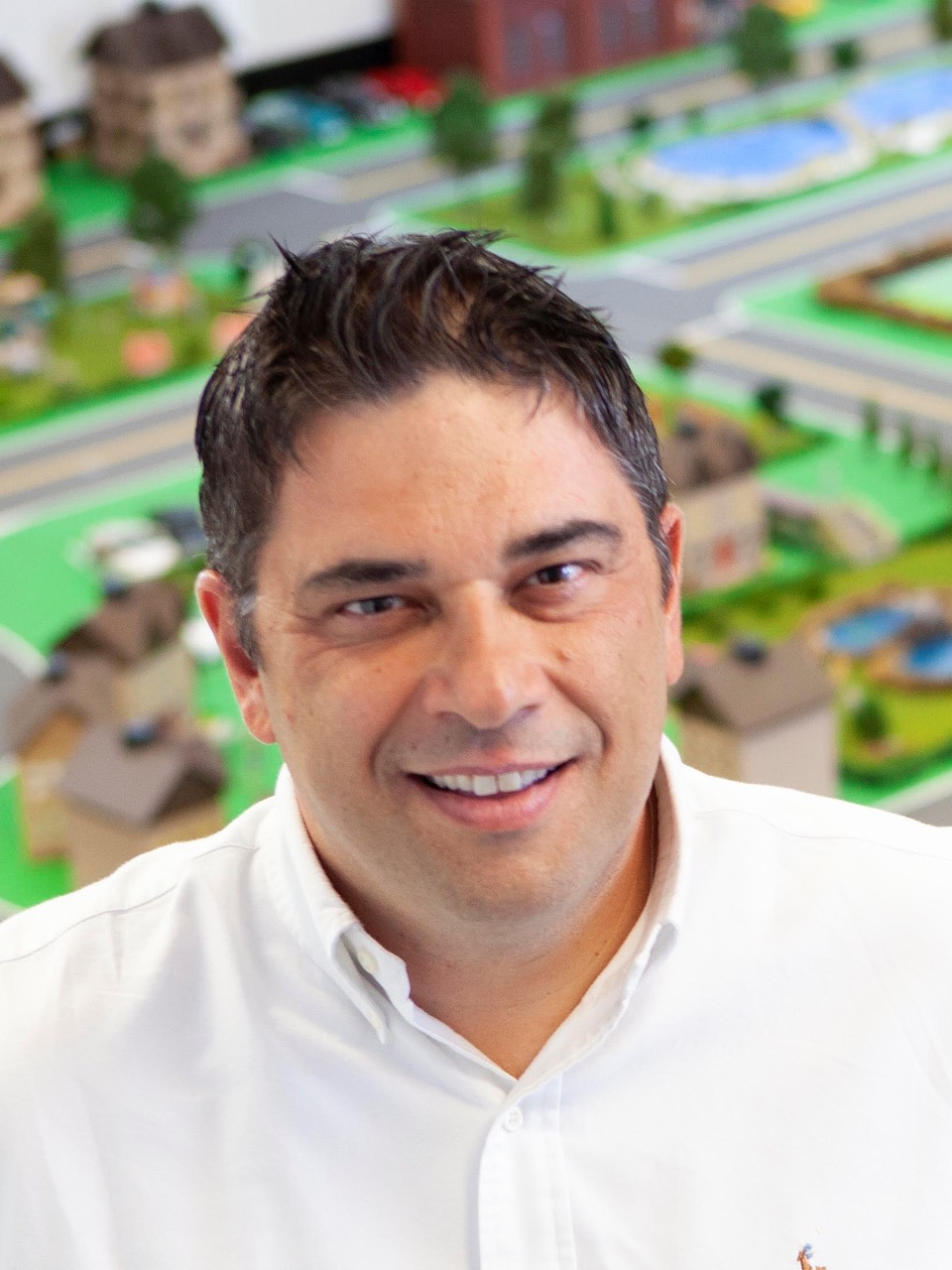}}]{Andreas A. Malikopoulos} (S’06–M’09–SM’17) received a Diploma in mechanical engineering from the National Technical University of Athens (NTUA), Greece, in 2000. He received M.S. and Ph.D. degrees in mechanical engineering at the University of Michigan, Ann Arbor, Michigan, USA, in 2004 and 2008, respectively. He is a professor at the School of Civil and Environmental Engineering at Cornell University and the director of the Information and Decision Science (IDS) Laboratory. Prior to these appointments, he was the Terri Connor Kelly and John Kelly Career Development Professor in the Department of Mechanical Engineering (2017-2023) and the founding Director of the Sociotechnical Systems Center (2019- 2023) at the University of Delaware (UD). Before he joined UD, he was the Alvin M. Weinberg Fellow (2010-2017) in the Energy \& Transportation Science Division at Oak Ridge National Laboratory (ORNL), the Deputy Director of the Urban Dynamics Institute (2014-2017) at ORNL, and a Senior Researcher in General Motors Global Research \& Development (2008-2010). His research spans several fields, including analysis, optimization, and control of cyber-physical systems (CPS); decentralized stochastic systems; stochastic scheduling and resource allocation; and learning in complex systems. His research aims to develop theories and data-driven system approaches at the intersection of learning and control for making CPS able to realize their optimal operation while interacting with their environment. He has been an Associate Editor of the IEEE Transactions on Intelligent Vehicles and IEEE Transactions on Intelligent Transportation Systems from 2017 through 2020. He is currently an Associate Editor of Automatica and IEEE Transactions on Automatic Control, and a Senior Editor of IEEE Transactions on Intelligent Transportation Systems. He is a member of SIAM, AAAS, and a Fellow of the ASME.
\end{IEEEbiography}

% \vfill

\end{document}